\documentclass[]{article}

\usepackage[T1]{fontenc}
\usepackage{lmodern}  
\usepackage[utf8]{inputenc}

\usepackage[
    top=1in,
    bottom=1in,
    left=0.9in,
    right=0.9in,
    includeheadfoot,
    headheight=13.6pt
]{geometry}

\usepackage[utf8]{inputenc}
\usepackage[T1]{fontenc}
\usepackage{amsmath}
\usepackage{amssymb}
\usepackage{graphicx}
\usepackage{hyperref}
\usepackage{authblk} 
\usepackage{url}
\usepackage{xspace}
\usepackage{booktabs}

\usepackage{bm} 
\usepackage{url}
\usepackage{amsmath}
\usepackage{amsfonts}
\usepackage{amsthm}
\usepackage{algorithmic}
\usepackage{graphicx}
\usepackage{textcomp}
\usepackage{xcolor}
\usepackage{soul} 
\usepackage{multirow}
\usepackage{array}
\usepackage{booktabs}

\def\BibTeX{{\rm B\kern-.05em{\sc i\kern-.025em b}\kern-.08em
    T\kern-.1667em\lower.7ex\hbox{E}\kern-.125emX}}

\usepackage{hyperref}
\usepackage[colorinlistoftodos]{todonotes}

\usepackage[ruled, noend]{algorithm2e}

\usepackage{tikz}

\usepackage{microtype}

\usepackage{relsize}

\usepackage{lipsum}

\newcommand{\nop}[1]{}

\newcommand{\set}[1]{\{#1\}}                    
\newcommand{\setof}[2]{\{{#1}\mid{#2}\}}        
\newcommand{\pr}{\mathop{\textnormal{Pr}}}    
\newcommand{\dom}{\textsf{Dom}}

\newcommand{\degree}{\texttt{deg}}

\newcommand{\calC}{\mathcal C}
\newcommand{\calX}{\mathcal X}

\newcommand{\defeq}{\stackrel{\text{def}}{=}}

\newcommand{\R}{\mathbb R} 
\newcommand{\Rp}{{\mathbb R}_{\tiny +}} 

\newcommand{\ce}{\textsc{CE}\xspace}
\newcommand{\pce}{\textsc{PCE}\xspace}
\newcommand{\est}{\textsc{Est}}

\newcommand{\attrs}{\texttt{Attrs}}
\newcommand{\vars}{\texttt{Vars}}

\newcommand{\lp}[1]{||#1||}

\newcommand{\system}{\textsc{LpBound}\xspace}
\newcommand{\psql}{\textsc{Postgres}\xspace}
\newcommand{\duckdb}{\textsc{DuckDB}\xspace}
\newcommand{\dbx}{\textsc{DbX}\xspace}
\newcommand{\safebound}{\textsc{SafeBound}\xspace}
\newcommand{\neurocard}{\textsc{NeuroCard}\xspace}
\newcommand{\bayescard}{\textsc{BayesCard}\xspace}
\newcommand{\deepdb}{\textsc{DeepDB}\xspace}
\newcommand{\flatcard}{\textsc{Flat}\xspace}
\newcommand{\factorjoin}{\textsc{FactorJoin}\xspace}

\newcommand{\groupby}{\texttt{group-by}\xspace}
\newcommand{\openclosed}[1]{(#1]}

\newcommand{\maxdegree}{\texttt{max-degree}\xspace}

\newcommand{\lpbase}{$\texttt{LP}_{\text{base}}$\xspace}
\newcommand{\lptd}{$\texttt{LP}_{\text{TD}}$\xspace}
\newcommand{\lptdb}{$\texttt{LP}_{\text{Berge}}$\xspace}
\newcommand{\lpflow}{$\texttt{LP}_{\text{flow}}$\xspace}
\newcommand{\nodes}{\text{Nodes}}
\newcommand{\edges}{\text{Edges}}

\definecolor{light-gray}{gray}{0.7.2}
\definecolor{goodgreen}{rgb}{0.1, 0.5, 0.1}
\definecolor{burntorange}{rgb}{0.8, 0.33, 0.0}
\definecolor{BlueViolet}{HTML}{8A2BE2}

\newcommand{\metarev}[1]{{\color{black}#1}}
\newcommand{\revone}[1]{{\color{black}#1}}
\newcommand{\revtwo}[1]{{\color{black}#1}}
\newcommand{\revthree}[1]{{\color{black}#1}}

\newtheorem{theorem}{Theorem}[section]
\newtheorem{lemma}[theorem]{Lemma}
\newtheorem{proposition}[theorem]{Proposition}

\theoremstyle{definition}
\newtheorem{definition}[theorem]{Definition}
\newtheorem{example}[theorem]{Example}

\title{\system: Pessimistic Cardinality Estimation using
  $\ell_p$-Norms of Degree Sequences}

\author[1]{Haozhe Zhang}
\author[1]{Christoph Mayer}
\author[2]{Mahmoud Abo Khamis}
\author[1]{Dan Olteanu}
\author[3]{Dan Suciu}

\affil[1]{University of Z\"urich, Switzerland}
\affil[2]{RelationalAI, United States}
\affil[3]{University of Washington, United States}

\date{}

\begin{document}

\maketitle

\begin{abstract}
  Cardinality estimation is the problem of estimating the size of the
  output of a query, without actually evaluating the query. The
  cardinality estimator is a critical piece of a query optimizer, and
  is often the main culprit when the optimizer chooses a poor plan.

  This paper introduces \system, a ``pessimistic'' cardinality
  estimator for multijoin queries (acyclic or cyclic) with 
  selection predicates and group-by clauses. \system computes a
  guaranteed upper bound on the size of the query output\nop{ that is very
  close to the true output size} using simple
  statistics on the input relations, consisting of $\ell_p$-norms of
  degree sequences.  The bound is the optimal solution of a linear
  program whose constraints encode data statistics and Shannon
  inequalities. We introduce two optimizations that exploit the
  structure of the query in order to speed up the estimation time and
  make \system practical.

  We experimentally evaluate \system against a range of traditional,
  pessimistic, and machine learning-based estimators on the JOB,
  STATS, and subgraph matching benchmarks. Our main finding is that
  \system can be orders of magnitude more accurate than traditional
  estimators used in mainstream open-source and commercial database
  systems. Yet it has comparable low estimation time and space
  requirements. When injected the estimates of \system, \psql derives
  query plans at least as good as those derived using the true
  cardinalities.
\end{abstract}

\section*{Keywords}
Cardinality Estimation, Degree Sequence, Lp-norms

\setcounter{page}{1}

\section{Introduction}
\label{sec:intro}

The \emph{Cardinality Estimation} problem, or \ce for short, is 
to estimate the output size of a query using only simple, precomputed
statistics on the database. \ce is one of the oldest and most
important problems in databases and data management.  It is used as the
primary metric guiding cost-based query optimization, for making
decisions about every aspect of query execution, ranging from broad
logical optimizations like the join order, to deciding the number of
servers to distribute the data over, and to detailed physical
optimizations, like the use of bitmap filters and memory allocation
for hash tables.

Unfortunately, \ce is notoriously difficult, and this affects
significantly the performance of data management systems.  Current
systems use density-based estimators, which were pioneered by System
R~\cite{DBLP:conf/sigmod/SelingerACLP79}.  They make drastic
simplifying assumptions (uniformity, independence, containment of
values, and preservation of values), and when the query has many joins
and many predicates, then they tend to have large errors, leading to
poor decisions by the downstream system; for example, the independence
assumption often leads to major
underestimation~\cite{DBLP:journals/pvldb/LeisGMBK015}.  Density-based
\ce also has limited support for queries with \groupby: most existing
systems yield poor estimates for the number of distinct
groups~\cite{DBLP:conf/cidr/Freitag019}. Yet the main problem with
traditional \ce is that it does not come with any theoretical
guarantees about its estimate: it may under-, or over-estimate, by a
little or by a lot, without any warning.  Several studies have shown
repeatedly that errors in the cardinality estimator can significantly
degrade the performance of most advanced database
systems~\cite{DBLP:journals/pvldb/LeisGMBK015,DBLP:journals/pvldb/LeeDNC23}.
To escape the simplifying assumptions of density-based CE, several
estimators were put forward that learn a model of the underlying
distribution in the database,
e.g.,~\cite{deepdb,bayescard,neurocard,flat}. This is a promising line
of work, yet as previously reported (and shown in our
experiments), their deployability is poor~\cite{FactorJoin:SIGMOD23} as
they lack explainability, have very slow training time, large model
size, and are difficult to transfer with comparable accuracy from one
workload to new workloads. One reason for this is that they need to
de-normalize the joined relations and add up to exponentially many
extra columns to represent new features.  Cardinality estimation
thus remains one of the major open challenges in data management.

In this paper, we introduce \system, a cardinality estimator 
that offers a one-sided guarantee: the true cardinality is guaranteed
to be below that returned by \system.  That is, \system returns
a guaranteed \emph{upper bound} on the size of the query  output.
Moreover, \system can explain the computed upper bound in terms of a
simple inequality, called a \emph{q-inequality}.  This one-sided
guarantee can be of use in many applications, for example it can
guarantee that a query does not run out of memory, or it can put an
upper bound on the number of servers required to distribute the output
data.  The challenge with this approach is to not overestimate too
much.  In other words, we want to reduce this upper bound as much as
possible, while still maintaining the theoretical one-sided guarantee.
To achieve that, we introduce novel statistics on the database, and
demonstrate that they lead to strictly improved upper bounds.  As an
extra bonus, \system applies equally well to \groupby queries.

There have been a small number of implementations that compute upper
bounds on the cardinality, commonly called \emph{pessimistic
  cardinality estimation}, or \pce for
short~\cite{DBLP:conf/sigmod/CaiBS19,DBLP:journals/pvldb/ChenHWSS22,DBLP:journals/tods/MhedhbiKS21}.
However, these systems were limited because they used only two types
of statistics on the input data: relation cardinalities, $|R|$, and
maximum degree (a.k.a.  maximum frequency) of an attribute $R.X$: if
the values of $R.X$ are $x_1, \ldots, x_N$, then the maximum degree is
$\max_i |\sigma_{X=x_i}(R)|$.  By using only limited input statistics,
these first-generation \pce systems led to significant overestimates,
and had worse accuracy than traditional, density-based \ce systems.

\begin{figure}[t]
\begin{minipage}{0.48\textwidth}
    \includegraphics[width=\textwidth,keepaspectratio]{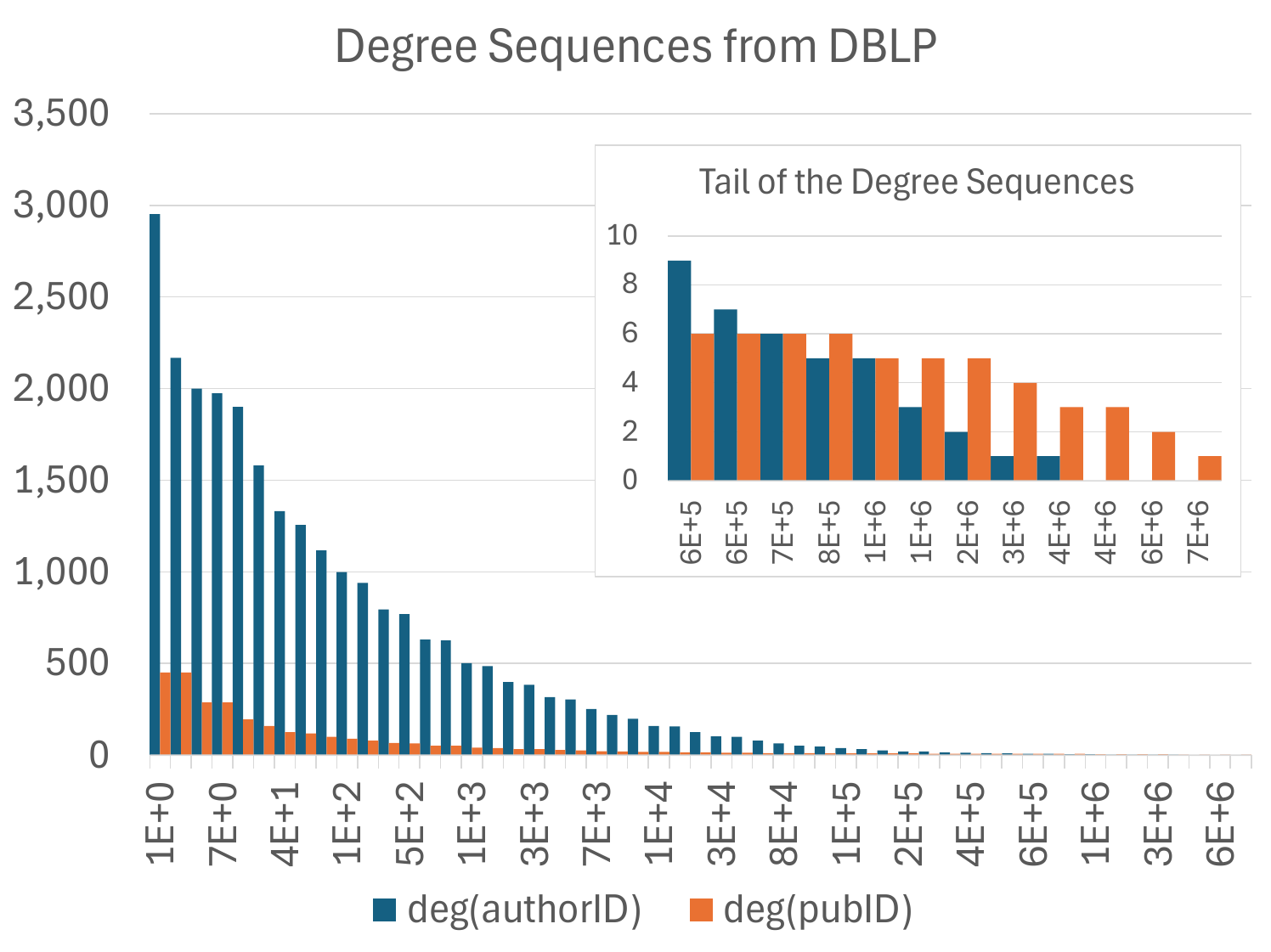}
    \caption{The \texttt{Authors}-\texttt{Publication} relationship in 
      DBLP (in 2023) with $24\cdot 10^6$ records pairing $3.6\cdot 10^6$ 
      authors and $7.1\cdot 10^6$ publications. The figure shows the 
      degree sequences $\degree(\texttt{authorID})$ and 
      $\degree(\texttt{pubID})$: The latter starts with a lower maximum degree, but 
      has a longer tail (see inset). The author with rank 1 is 
      \texttt{H. Vincent Poor} ($2951$ publications) and the publication 
      with rank 1 is~\cite{DBLP:journals/tmlr/SrivastavaRRSAF23} (with 
      $450$ authors).}
    \label{fig:ds}
\end{minipage}
\hfill
\begin{minipage}{0.48\textwidth}
    \includegraphics[width=\textwidth,keepaspectratio]{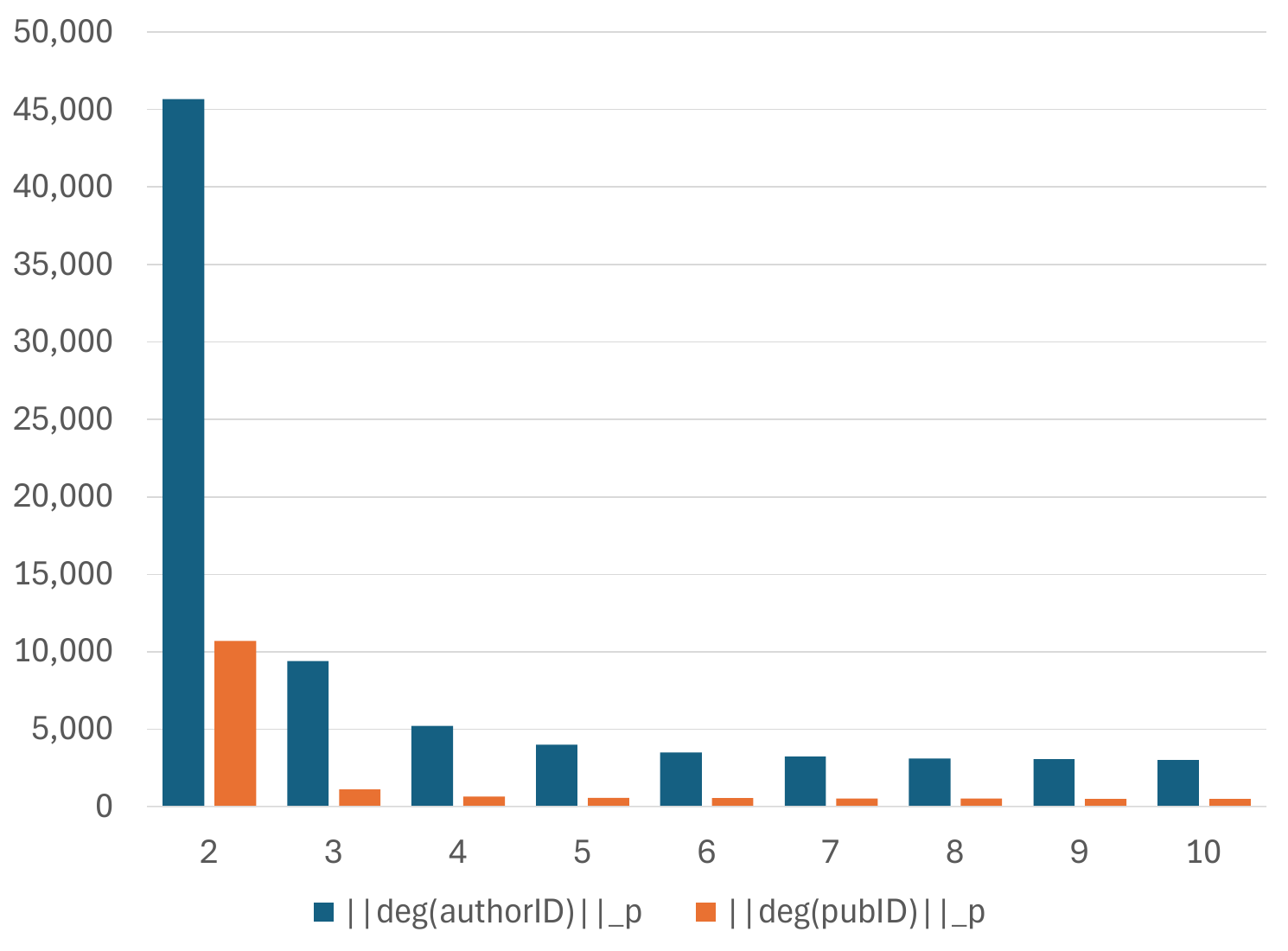}
    \caption{Instead of storing the two degree sequences, \system stores 
      only some of their $\ell_p$-norms, for example for $p\in \set{1,\ldots, 10,\infty}$ 
      shown here. We do not show $\ell_1$ (it is equal to the cardinality 
      \revthree{$24\cdot 10^6$}) and $\ell_\infty$ (it is the maximum degree, 
      $2951$ or $450$ respectively). When $p$ ranges from $1$ to $\infty$, the 
      $\ell_p$-norm ranges from the relation's cardinality to the maximum 
      degree of any value.}
    \label{fig:lp}
\end{minipage}
\end{figure}

To achieve better upper bounds, we use significantly richer statistics
on the input database.  Concretely, we use the {\em $\ell_p$-norms of
  degree sequences} as inputs to \system.  The \emph{degree sequence}
of an attribute $R.X$ is the sequence
$\degree_R(X) = (d_1, d_2, \ldots, d_N)$, sorted in decreasing order,
where $d_i$ is the frequency of the value $x_i$.  The $\ell_p$-norm is
$\left(\sum_i d_i^p\right)^{1/p}$.  The $\ell_p$-norms of degree
sequences are related to \emph{frequency moments}~\cite{DBLP:conf/stoc/AlonMS96}: 
The $p$'th frequency moment is $\sum_i d_i^p$. These are 
commonly used in statistics and machine learning, since they
capture important information about the data distribution.  This
information can be very useful for cardinality estimation too. 
It is also practical as it can be computed and maintained efficiently~\cite{DBLP:conf/stoc/AlonMS96}. 
Yet, to the best of our knowledge, the $\ell_p$-norms (or frequency moments)
have not been used before for cardinality estimation.  \system is, to
the best of our knowledge, the first to use them for cardinality
estimation.
Fig.~\ref{fig:ds} shows the degree sequences of the
Author-Publication relationship in the DBLP database, and
Fig.~\ref{fig:lp} shows some of their $\ell_p$-norms.

\system takes as input a query with equality joins,
equality and range predicates, and group-by clause, and computes an upper
bound on the query output size, by using precomputed $\ell_p$-norms on the input
database. \system offers a strong, theoretical guarantee: for any database that satisfies the
given statistics, the query output size is guaranteed to be below the
bound returned by \system.  The bound is \emph{tight}, in the sense
that, if all we know about the input database are the given
statistics, then there exists a worst-case input database with these
statistics on which the query output is as large as the bound
returned by \system.  Finally, \system is able to explain the upper
bound, in terms of a simple \emph{q-inequality} relating the output
size to the input statistics.

\paragraph{Contributions} In this paper we make four main contributions.  

{\em 1. We introduce \system, a PCE that uses $\ell_p$-norms of input relations (Sec.~\ref{sec:lpbound})}. 
\system is a principled framework to compute the upper bound, based on
information theory, building upon, and expanding a long line of
theoretical
results~\cite{DBLP:journals/siamcomp/AtseriasGM13,DBLP:journals/jacm/GottlobLVV12,DBLP:conf/pods/KhamisNS16,DBLP:conf/pods/Khamis0S17,DBLP:journals/pacmmod/KhamisNOS24}.
We show how to extend previous results to 
accommodate \groupby queries. \system works for both cyclic and
acyclic queries, and therefore can be used as an estimator both for
traditional SQL workloads, which tend to be acyclic, and for graph
pattern matching or SparQL queries, which tend to be cyclic.  

{\em 2. We describe how to use most common values and histograms to extend \system to conjunctions and disjunctions of equality and range predicates (Sec.~\ref{sec:histograms}).}
To support predicates, \system uses data structures that are very similar to those used by SQL engines, and therefore \system could be easily incorporated in those systems.

{\em 3. We introduce two optimization techniques for computing the upper bound, which
run in polynomial time in the size of the query and the number of
available statistics (Sec.~\ref{sec:algorithm}).} One works for acyclic queries only, while the other works for arbitrary conjunctive queries albeit on one-column degree sequences.
These techniques are essential for the practicality of \system, as  cardinality estimation is often invoked thousands of times during query optimization, and it must run in times
measured in milliseconds.

{\em 4. We conduct an extensive experimental evaluation of \system on real and
  synthetic workloads (Sec.~\ref{sec:experiments}).}
  \system can be orders of magnitude more accurate than traditional
  estimators used in mainstream open-source and commercial database
  systems. Yet it has low estimation time and space
  requirements to remain practical. 
  When injected the estimates of \system, \psql derives
  query plans at least as good as those derived using the true
  cardinalities.

\paragraph{Related Work}
Our paper builds on a long line of theoretical
results that proved upper bounds on the size of the query output.  
The first such result appeared in a landmark paper by Atserias, Grohe, and
Marx~\cite{DBLP:journals/siamcomp/AtseriasGM13}, which proved an upper
bound in terms of the cardinalities of the input
relations, known today as the \emph{AGM Bound}.  The AGM bound
is not practical for real SQL workloads, which consist almost
exclusively of acyclic queries, where the AGM bound is too large.
For example, the AGM bound of a 2-way join is
$|R \Join S| \leq |R| \cdot |S|$.  The AGM bound was extended to
account for functional dependencies~\cite{DBLP:journals/jacm/GottlobLVV12,DBLP:conf/pods/KhamisNS16},
and further extended to use the maximum degrees of
attributes~\cite{DBLP:conf/pods/Khamis0S17}: we refer to the
latter as the \emph{\maxdegree bound}.  This line of work relies on information theory.  A simplified version of the \maxdegree bound was incorporated into two pessimistic cardinality estimators~\cite{DBLP:conf/sigmod/CaiBS19,DBLP:conf/cidr/HertzschuchHHL21}.
However, cardinalities and maximum degrees alone are still too limited to infer useful upper bounds for acyclic queries.  For example, the \maxdegree bound of a 2-way join is
$|R \Join_{X=Y} S| \leq \min(a\cdot|S|, |R|\cdot b)$, where $a$ is the maximum degree of $R.X$ and $b$ is the maximum degree of $S.Y$.  Real data is often skewed and the maximum degree is large (the maximum degrees are $2951$ and $450$ in Fig.~\ref{fig:ds}), and this led to large overestimates for more complex queries.

The first system to use degree sequences for pessimistic
cardinality estimation was
\safebound~\cite{SafeBound:SIGMOD23,DBLP:conf/icdt/DeedsSBC23}.
Since the degree sequences are often too large, 
\safebound uses a lossy compression of them. It relies solely on combinatorics, it is limited to Berge-acyclic queries (see Sec.~\ref{sec:background}), and it does not support \groupby. \system can be seen as a significant extension of \safebound.  By using information theory instead of combinatorics, \system computes the bound using $\ell_p$-norms, without requiring the degree
sequences, and also works for cyclic queries and queries
with \groupby.

\section{Background}
\label{sec:background}

\paragraph{Relations}
We write $\attrs(R)$ for the set of attributes of a relation $R$.  The domain of attributes $U \subseteq \attrs(R)$  is $\dom(R.U) \defeq \Pi_U(R)$.

\paragraph{Queries}
\system supports single-block SQL queries:
\begin{align*}
  & \texttt{SELECT } \text{[groupby-attrs]}\\
  & \texttt{FROM } R_1, R_2, \ldots, R_m \\
  & \texttt{WHERE } \text{[join-and-selection-predicates]}\\
  & \texttt{GROUP BY }\text{[groupby-attrs]}
\end{align*}
We ignore aggregates in \texttt{SELECT} because they do not affect the output cardinality.
We do not support sub-queries.  The predicates can be equality, range, and their conjunction and disjunction; 
\texttt{IN} and \texttt{LIKE} predicates can be supported with trivial
effort (see Sec.~\ref{sec:histograms}).
\nop{
For example:
\newline $\texttt{City IN ('Pittsburg','Edinburg') OR City LIKE '\%burgh\%'}$
}
\revone{Queries with bag semantics are also supported, by replacing
  them with a full query.  For
  example, the output of $\texttt{SELECT A FROM }\ldots$
  (without \texttt{GROUP BY}) is a bag that has the same size as
  the output of $\texttt{SELECT * FROM }\ldots$, which is a set.
  Throughout the paper we will assume that queries have set semantics.}
We will use the conjunctive query notation instead of SQL:
\begin{align}
  Q(V_0) = & R_1(V_1) \wedge R_2(V_2) \wedge \cdots \wedge R_m(V_m)\wedge\text{[predicates]} \label{eq:cq}
\end{align}
where each $V_j\defeq \attrs(R_j)$ is a set of variables, and
$V_0 \subseteq V_1 \cup \cdots \cup V_m$ represents the \groupby
variables.  We denote by
$\vars(Q) \defeq V_1 \cup \cdots \cup V_m = \set{X_1, \ldots, X_n}$
the set of all variables in the query.  When $V_0=\set{X_1, \ldots, X_n}$, 
then we say that $Q$ is a \emph{full} conjunctive
query.  $Q$ is \emph{acyclic} if its relations $R_1, \ldots, R_m$ can
be placed on the nodes of a tree, such that, for every individual
variable $X_i$, the set of tree nodes that contain $X_i$ forms a
connected component.  $Q$ is called \emph{Berge-acyclic} if it is
acyclic and any two relations share at most one variable.  For
example, the 3-way join query $J_3$ is Berge-acyclic, while the
3-clique query $C_3$ is cyclic:
\begin{align}
  J_3(X,Y,Z,U) = & R(X,Y) \wedge S(Y,Z) \wedge T(Z, U) \label{eq:j3}\\
  C_3(X,Y,Z)   = & R(X,Y) \wedge S(Y,Z) \wedge T(Z,X) \label{eq:c3}
\end{align}

\paragraph{Degree Sequences}
Fix a relation instance $R$, and two
sets of variables $X, Y \subseteq \attrs(R)$.  The \emph{degree
  sequence} from $X$ to $Y$ in $R$ is the sequence
$\degree_R(Y|X) \defeq (d_1, d_2, \ldots, d_N)$ obtained as follows.
Compute the domain of $X$, $\dom(R.X)=\set{x_1, \ldots, x_N}$, denote
by $d_i = |\sigma_{X=x_i}(\Pi_{XY}(R))|$ the degree (or frequency) of
$x_i$, and sort the values in the domain $\dom(R.X)$ such that their
degrees are decreasing $d_1 \geq d_2 \geq \cdots \geq d_N$.  We call
$i$ the \emph{rank} of the element $x_i$.  Notice that
$\degree_R(Y|X)$ and $\degree_R(XY|X)$ are the same, where $XY$
denotes the union $X \cup Y$.  When $X=\emptyset$, then the degree
sequence has length 1, $\degree_R(Y|\emptyset)=(|\dom(R.Y)|)$.  When
the functional dependency $X \rightarrow Y$ holds (for example, if $X$
is a key), then $\degree_R(Y|X)=(1,1,\ldots,1)$.  When $|X|\leq 1$, then
we say that the degree sequence is \emph{simple}, and when
$XY =\attrs(R)$, then we say that the degree sequence is \emph{full}
and denote it by $\degree_R(*|X)$, or just $\degree_R(X)$ (we used
this notation in Fig.~\ref{fig:ds}).  In this paper we will consider
only simple degree sequences.
\revone{The degree sequence also
  applies to the case when $R$ is a bag, not necessarily a set.}
%
%
The $\ell_p$-norm of a sequence $\bm d = (d_1, d_2, \ldots)$ is
$\lp{\bm d}_p \defeq \left(\sum_i d_i^p\right)^{1/p}$, where
$p \in \openclosed{0,\infty}$.  When $p$ increases towards $\infty$,
$\lp{\bm d}_p$ decreases and converges to
$\lp{\bm d}_\infty \defeq \max_i d_i$, see Fig.~\ref{fig:lp}.

Fig.~\ref{fig:degree} illustrates some simple examples of degree
sequences: $\degree_R(YZ|X)$ is both simple and full, and we can write
it as $\degree_R(*|X)$ or just $\degree_R(X)$. The degree sequence
$\degree_R(Z|XY)$ is not simple.

\begin{figure}
  \centering
    \begin{align*}
      R=&\begin{array}[c]{|c|c|c|} \hline X&Y&Z \\ \hline
         1&a&\ldots \\
         1&b&\ldots \\
         1&b&\ldots \\
         2&a&\ldots \\
         2&b&\ldots \\
         3&b&\ldots \\
         3&c&\ldots \\
         4&d&\ldots \\ \hline
    \end{array}
&&
   \begin{array}[c]{ll}
     \degree_R(YZ|X)&=(3,2,2,1)\\
     \degree_R(Y|X) &= (2,2,2,1)\\
     \degree_R(Z|XY)&=(2,1,1,1,1,1,1)\\
     \degree_R(XYZ|\emptyset)&=(8)=(|R|)
   \end{array}
    \end{align*}
    \caption{Examples of Degree Sequences.}
  \label{fig:degree}
\end{figure}

\paragraph{Density-based \ce}
The traditional, density-based cardinality estimator~\cite{DBLP:books/daglib/0020812} is
limited to selections and joins.  It makes the assumptions mentioned
in the introduction and computes the estimate bottom-up on the query
plan, for example:
\begin{align*}
  \est(\sigma_{X=\text{value}}(R)) = & \frac{|R|}{|\dom(R.X)|} \\
  \est(R \Join_{X=Y} S) = & \frac{|R|\cdot|S|}{\max(|\dom(R.X)|,|\dom(S.Y)|)}
\end{align*}
The ratio $\frac{|R|}{|\dom(R.X)|}$ is the average degree,
$\texttt{Avg}(\degree_R(*|X))$.

Queries with \groupby are treated differently by different systems.
We describe briefly how they are handled by two open-source systems,
illustrating on the following \groupby query:
\begin{align}
    JG_3(X,U) = & R(X,Y) \wedge S(Y,Z) \wedge T(Z, U) \label{eq:jg3}
\end{align}
\duckdb ignores the \groupby clause, and estimates the size of $JG_3$
to be the same as that of the full join $J_3$ (Eq.~\eqref{eq:j3}).
\psql estimates it as the minimum between the full join, and the
product of the domains of the \groupby variables,
$|\dom(R.X)|\cdot |\dom(T.U)|$.

\paragraph{Theoretical Upper Bounds}
An \emph{upper bound} for a
conjunctive query $Q$ is a numerical value, which is computed in terms
of statistics on the input database,  such as the output size of the query is
is guaranteed to be below that bound.  The upper bound is \emph{tight}
if there exists a database instance, satisfying the statistics, such
that the query's output is as large as the bound.\footnote{Up to some
  small, query-dependent constant.}  The AGM
bound~\cite{DBLP:journals/siamcomp/AtseriasGM13} is a tight upper that
uses only the cardinalities $|R_1|, \ldots, |R_m|$; in other words, it
uses only the $\ell_1$-norms of full degree sequences.  A non-negative
sequence $w_1, \ldots, w_m$ is called a \emph{fractional edge cover}
of the query $Q$ in Eq.~\eqref{eq:cq} if every variable $X_i$ is
``covered'', meaning that $\sum_{j\in[m]: X_i \in V_j}w_j \geq 1$.  The AGM
bound states that
$|Q| \leq |R_1|^{w_1}\cdot|R_2|^{w_2}\cdots|R_m|^{w_m}$ for any
fractional edge cover.  It is useful for cyclic
queries like $C_3$ above (see Sec.~\ref{subsec:lpbound:full}), but for
acyclic queries it degenerates to a product of cardinalities, because
the optimal edge cover is integral.  For example, the AGM bound of the
3-way join in Eq.~\eqref{eq:j3} is $|J_3| \leq |R|\cdot |T|$, because
the optimal fractional edge cover is\footnote{Every fractional edge
  cover must satisfy $w_R\geq 1$ in order to cover $X$, and
  $w_T \geq 1$ to cover $U$; then $w_S$ can be arbitrary.  Therefore,
  $|R|^{w_R}|S|^{w_S}|T|^{w_T}\geq |R|\cdot|T|$.}
$w_R=1, w_S=0, w_T=1$.

The \maxdegree bound introduced in~\cite{DBLP:conf/pods/Khamis0S17}
generalizes the AGM bound by using both cardinalities and maximum
degrees; in other words, it uses both $\ell_1$ and $\ell_\infty$ norms
of degree sequences.  When restricted to acyclic queries, the
\maxdegree bound represents an improvement over the AGM, but it is
still less accurate than a density-based estimate.
%
%
For example, the bound for $J_3$ is the minimum of the following four
quantities (see also~\cite{DBLP:journals/pvldb/ChenHWSS22}):
\begin{align}
  |J_3|\leq & |R|\cdot |T| \nonumber \\
  |J_3|\leq & |R| \cdot \lp{\degree_S(Z|Y)}_\infty \cdot\lp{\degree_T(U|Z)}_\infty\label{eq:degree:bound:j3}\\
  |J_3|\leq & |S|\cdot \lp{\degree_R(X|Y)}_\infty \cdot  \lp{\degree_T(U|Z)}_\infty \nonumber \\
  |J_3| \leq &|T|\cdot \lp{\degree_S(Y|Z)}_\infty \cdot \lp{\degree_R(X|Y)}_\infty \nonumber 
\end{align}
For comparison, the traditional, density-based estimator for $J_3$ is:
{
\begin{align*}
  \est(J_3)=&\frac{|R|\cdot|S|\cdot|T|}{\max(|\dom(R.Y)|,|\dom(S.Y)|)\cdot\max(|\dom(S.Z)|,|\dom(T.Z)|)}
\end{align*}
}

When $|\dom(R.Y)|\leq |\dom(S.Y)|$ and $|\dom(S.Z)|\leq |\dom(T.Z)|$
then the estimator becomes
$|R|\cdot
\texttt{Avg}(\degree_S(Z|Y))\cdot\texttt{Avg}(\degree_T(U|Z))$, which
is the same as the \maxdegree bound in Eq.~\eqref{eq:degree:bound:j3}
with the maximum degree replaced by the average degree.

\safebound~\cite{DBLP:conf/icdt/DeedsSBC23,SafeBound:SIGMOD23}
uses simple, full degree sequences and computes a tight upper bound of
a Berge-acyclic, full conjunctive query.  For example, if
$\degree_R(*|X)=(a_1\geq a_2 \geq \cdots)$ and
$\degree_S(*|Y)=(b_1\geq b_2 \geq \cdots)$, then \safebound will infer
the following bound on a 2-way join:
$|R \Join_{X=Y} S| \leq \sum_i a_ib_i$.  When applied to the 3-way
join $J_3$ \safebound returns a much better bound than the degree
bound~\eqref{eq:degree:bound:j3}, but that bound is not described by a
closed-form formula; it is only given by an algorithm.  The
limitations of \safebound are its lack of explainability, its
restriction to Berge-acyclic queries, and its reliance on 
compression heuristics for the degree sequences.

\paragraph{Information Theory} Let $X$ be a finite random
variable, with outcomes $x_1, \ldots, x_N$, and probability function
$\pr$.  Its \emph{entropy} is:
\begin{align*}
  h(X) \defeq & - \sum_{i=1,N} \pr(x_i) \log \pr(x_i)
\end{align*}
where $\log$ is in base 2.  It holds that $0 \leq h(X) \leq \log N$,
and $h(X)=\log N$ iff $\pr$ is uniform, i.e.,
$\pr(x_1)=\cdots=\pr(x_N)=1/N$.

Let $X_1,\ldots, X_n$ be $n$ finite, jointly distributed random
variables.  They can be described by a finite relation
$R(X_1, \ldots, X_n)$, representing their support, and a probability
function s.t. for each tuple $t \in R$, $\pr(t)\geq 0$ and
$\sum_{t \in R}\pr(t)=1$.  For every subset $U$ of variables, $h(U)$
denotes the entropy of the marginal distribution of the random variables
in $U$.  For example, we have $h(X_1X_3)$, $h(X_2X_4X_5)$, etc.  This
defines a vector $h$ with $2^n$ dimensions, which is called an
\emph{entropic vector}.  The \emph{conditional entropy} is defined as
\begin{align}
  h(V|U) \defeq & h(UV) - h(U) \label{eq:chain}
\end{align}
The following hold for all subsets of variables
$U, V \subseteq \attrs(R)$:
\begin{align}
  h(V) \leq & \log|\dom(R.V)| & h(V|U) \leq & \log\lp{\degree_R(V|U)}_\infty \label{eq:h:log}
\end{align}
Every entropic vector $\bm h$ satisfies the \emph{basic Shannon
  inequalities}:
\begin{align}
&&  h(\emptyset) =&\ 0\nonumber\\
&\text{Monotonicity:}&  h(U\cup V) \geq &\  h(U)\label{eq:shannon:monotonicity}\\
&\text{Submodularity:}&  h(U)+h(V)\geq &\  h(U\cup V)+h(U\cap V)\label{eq:shannon:submodularity}
\end{align}
Every vector $\bm h$ that satisfies the basic Shannon inequalities is
called a \emph{polymatroid}.  Every entropic vector is a polymatroid,
but the converse is not
true~\cite{zhang1998characterization,Yeung:2008:ITN:1457455}.
%

\section{The \system Cardinality Estimator}
\label{sec:lpbound}

Our system, \system, is a significant extension of previous upper
bound estimators, in that it computes a tight upper bound of the query $Q$ by
using $\ell_p$-norms of simple degree sequences.  \system can explain
its upper bound in terms of a simple inequality, called a
\emph{q-inequality}.  We introduce \system gradually, by first
describing the q-inequalities, and showing later how to compute the
optimal bound.  Throughout this section, we assume that the query has
no predicates: we discuss predicates in Sec.~\ref{sec:histograms}.

\subsection{Q-Inequalities for Full Queries}

\label{subsec:lpbound:full}

For upper bounds on a full conjunctive query in terms of
$\ell_p$-norms, we use inequalities described
in~\cite{DBLP:journals/pacmmod/KhamisNOS24}.  As a simple warmup,
consider the 2-way join, which we write as:
\begin{align}
  J_2(X,Y,Z) = & R(X,Y) \wedge S(Y,Z) \label{eq:j2}
\end{align}
\system uses the following q-inequalities:
\begin{align}
  |J_2| \leq & |R|\cdot|S| \\
  |J_2| \leq & |R|\cdot \lp{\degree_S(*|Y)}_\infty \\
  |J_2| \leq & \lp{\degree_R(*|Y)}_\infty \cdot |S| \\
  |J_2| \leq & \lp{\degree_R(*|Y)}_2 \cdot \lp{\degree_S(*|Y)}_2
\end{align}
The first bound is the AGM bound; the next two are the \maxdegree
bound, and are always lower (i.e. better) than the AGM bound.  The
last bound is new, and follows from the Cauchy-Schwartz inequality.
\system always returns the smallest value of all q-inequalities.  It
does not need to enumerate all of them; instead it computes the bound
differently (explained below in Sec.~\ref{subsec:basic:algorithm}),
then returns as explanation the single q-inequality that produces that
bound.

For the 3-way join $J_3$ from Eq.~\eqref{eq:j3}, \system uses many more q-inequalities.  It
includes all those considered by the \maxdegree bound
(Eq.~\eqref{eq:degree:bound:j3}) and many more.  We show here only two
q-inequalities:

{
\begin{align}
  |J_3| \leq & \lp{\degree_R(X|Y)}_2 \cdot |S|^{1/2} \cdot\lp{\degree_T(U|Z)}_2 \nonumber\\
  |J_3| \leq & |R|^{1/3}\cdot \lp{\degree_R(X|Y)}_2^{2/3} \cdot \lp{\degree_S(Z|Y)}_{2}^{2/3}\cdot\lp{\degree_T(U|Z)}_3 \label{eq:pce:j3:p3:r}
\end{align}
}

To the best of our knowledge, such inequalities have not been used
previously in cardinality estimation.  We prove~\eqref{eq:pce:j3:p3:r}
in Sec.~\ref{subsec:sec:lpbound:proofs}.  \system also improves significantly the bounds of cyclic
queries, for example it considers these q-inequalities for the
3-clique $C_3$:
\begin{align}
  |C_3| \leq & \left(|R|\cdot |S| \cdot |T|\right)^{1/2}\nonumber \\
  |C_3| \leq & \left(\lp{\degree_R(Y|X)}_2^2\cdot\lp{\degree_S(Z|Y)}_2^2\cdot\lp{\degree_T(X|Z)}_2^2\right)^{1/3}\nonumber \\
  |C_3| \leq & \left(\lp{\degree_R(Y|X)}_3^3\cdot\lp{\degree_S(Y|Z)}_3^3\cdot|T|^5\right)^{1/6}\label{eq:pce:t:3} 
\end{align}
The first is the AGM bound corresponding to the fractional edge cover
$w_R=w_S=w_T=\frac{1}{2}$.  The other two are novel and surprising.

\subsection{\system for \groupby Queries}
\label{sec:lpbound-groupby-queries}

Similar q-inequalities hold for queries with \groupby.  We 
illustrate here for the query $JG_3$ in Eq.~\eqref{eq:jg3}.

Every q-inequality that holds for the full conjunctive query also
holds for the \groupby query, in other words $|JG_3|\leq |J_3|$, and
all upper bounds for $J_3$ also apply to $JG_3$; this is used by \duckdb.

Further q-inequalities can be obtained by dropping variables that do not occur in \groupby, as done in \psql.  For example, we can drop the variables $Y,Z$ from $JG_3$
and obtain the query:
\begin{align*}
  JG_3'(X,U) = & R'(X) \wedge T'(U)
\end{align*}
for which we can infer:
\begin{align*}
  |JG_3|\leq& |JG_3'| \leq |R'|\cdot|T'| = |\dom(R.X)|\cdot|\dom(T.U)|
\end{align*}

However, \system uses many more q-inequalities, which are not
necessarily derived using the two heuristics above.  For example,
consider the following star-join with \groupby:
\begin{align*}
  \text{StarG}(X_1,X_2) = & R_1(X_1,Z)\wedge R_2(X_2,Z)\wedge S(Y,Z)
\end{align*}
\system infers (among others) the following inequality:
\begin{align}
  |\text{StarG}| \leq & |S|^{1/3}\cdot \lp{\degree_{R_1}(X_1|Z)}_3\cdot\lp{\degree_{R_2}(X_2|Z)}_3\label{eq:sg:bound}
\end{align}
This q-inequality does not hold for the full conjunctive
query\footnote{Proof: consider the instance $R_1=R_2=\set{(1,1)}$,
  $S=\set{(1,1),(2,1),\ldots,(N,1)}$.  The full join returns an output
  of size $N$, while the RHS of~\eqref{eq:sg:bound} is $N^{2/3}$.} and
it involves all query variables.  We prove~\eqref{eq:sg:bound} below.

\subsection{Proofs of Q-Inequalities}
\label{subsec:sec:lpbound:proofs}

Consider $n$ finite random variables $X_1, \ldots, X_n$, and let their
set of outcomes be the relation $R(X_1, \ldots, X_n)$ (see
Sec.~\ref{sec:background}).  Then, for any subsets of variables
$U, V \subseteq \attrs(R)$ and any $p \in \openclosed{0,\infty}$, the
following holds~\cite{DBLP:journals/pacmmod/KhamisNOS24}:
\begin{align}
  \frac{1}{p}h(U)+h(V|U) \leq & \log \lp{\degree_R(V|U)}_p \label{eq:h:p}
\end{align}
Inequalities~\eqref{eq:h:log} are special cases of \eqref{eq:h:p}, where 
$p=1$ or $p=\infty$.

Inequality~\eqref{eq:h:p} is very important.  It connects an
information-theoretic term in the LHS with a statistics on the input
database in the RHS.  All q-inequalities inferred by \system follow
from~\eqref{eq:h:p} and the basic Shannon inequalities.  We illustrate
with two examples.

First, we prove the q-inequality~\eqref{eq:pce:j3:p3:r} for $J_3$.
Assume three input relations $R(X,Y)$, $S(Y,Z)$, $T(Z,U)$, and denote
by $N \defeq |J_3|$ the size of the query's output. Consider the
uniform probability distribution with outcomes $J_3$: every tuple
$t = (x,y,z,u)\in J_3$ has the same probability, $\pr(t)= 1/N$.
Therefore, their entropy is $h(XYZU)=\log |J_3|$ (by uniformity),
and~\eqref{eq:pce:j3:p3:r} follows from:

{
  \begin{align*}
    \log & |R| +  2 \log \lp{\degree_R(X|Y)}_2 +  2 \log  \lp{\degree_S(Z|Y)}_{2} + 3 \log \lp{\degree_T(U|Z)}_3 \geq \\
    \geq & h(XY) + 2\left(\frac{1}{2}h(Y)+h(X|Y)\right)+2\left(\frac{1}{2}h(Y)+h(Z|Y)\right)+3\left(\frac{1}{3}h(Z)+h(U|Z)\right)
          \text{\hspace{2mm}by~\eqref{eq:h:p}}\\
    = & h(XY)+\left(h(XY)+h(X|Y)\right)+\left(h(YZ)+h(Z|Y)\right)+\left(h(UZ)+2h(U|Z)\right)
         \text{\hspace{6mm}by~\eqref{eq:chain}}\\
    = & \left(h(XY) + h(Z|Y)+ h(U|Z)\right) + \left(h(XY)+h(UZ)\right)+\left(h(X|Y)+h(YZ)+h(U|Z)\right)  \\
    \geq & \left(h(XY) + h(Z|XY)+ h(U|XYZ)\right)  + h(XYZU) + \left(h(X|YZ)+h(YZ)+h(U|XYZ)\right)\\
         &\text{\hspace{60mm}by submodularity}\\
    = & 3 h(XYZU) = 3 \log |J|
  \end{align*}
}

The first inequality is an application of~\eqref{eq:h:p}.  The second
inequality uses submodularity, for example $h(Z|Y) \geq h(Z|XY)$
follows from $h(YZ)-h(Y) \geq h(XYZ)-h(XY)$, or
$h(XY)+h(YZ)\geq h(XYZ)+h(Y)$.

Second, we prove the q-inequality~\eqref{eq:sg:bound}.  The setup is
similar: assume some input relation instances\\
$R_1(X_1,Z), R_2(X_2,Z), S(Y,Z)$, let $\text{StarG}(X_1,X_2)$ be the output of
the query, and denote by $N \defeq |\text{StarG}|$.  Define $\text{Star}(X_1,X_2,Y,Z)$ to
be the result of the full join.  We need a probability distribution on
$\text{Star}$ whose marginal on $X_1,X_2$ is uniform.  There are many ways to
define such a distribution, we consider the following.  Order the
tuples in $\text{Star}$ arbitrarily; then, for each tuple $t = (x_1,x_2,y,z)$,
if there exists some earlier tuple with the same values $x_1,x_2$ then
set $\pr(t)=0$, otherwise set $\pr(t)=1/N$.  At this point, we continue
similarly to the previous example:

\nop{
\begin{align*}
  \frac{2}{3}\log & |S| + \log \lp{\degree_{R_1}(X_1|Z)}_3+\log \lp{\degree_{R_2}(X_2|Z)}_3\geq\\
  \geq & \frac{2}{3}h(YZ)+\frac{1}{3}h(Z)+h(X_1|Z)+\frac{1}{3}h(Z)+h(X_2|Z) \\
  = & \frac{1}{3}\biggl(2\bigl(h(YZ)+h(X_1|Z)+h(X_2|Z)\bigr) + \bigl(h(X_1Z)+h(X_2Z)\bigr)\biggr)\\
  \geq  & \frac{1}{3}\left(2h(X_1X_2YZ) + h(X_1X_2Z)\right)\text{\hspace{20mm}submodularity}\\
  \geq & h(X_1X_2)=\log|\text{StarG}| \text{\hspace{25mm}monotonicity}
\end{align*}
}

\begin{align*}
  \frac{1}{3}\log & |S| + \log \lp{\degree_{R_1}(X_1|Z)}_3+\log \lp{\degree_{R_2}(X_2|Z)}_3\geq\\
  \geq & \frac{1}{3}h(YZ)+\frac{1}{3}h(Z)+h(X_1|Z)+\frac{1}{3}h(Z)+h(X_2|Z) \\
  \geq & \frac{1}{3}h(Z)+\frac{1}{3}h(Z)+h(X_1|Z)+\frac{1}{3}h(Z)+h(X_2|Z) \\
  = & h(Z) + h(X_1|Z)+ h(X_2|Z) = h(X_1Z) + h(X_2|Z) \\ 
  \geq & h(X_1Z) + h(X_2|X_1Z)  = h(X_1X_2Z) 
  \geq h(X_1X_2)=\log|\text{StarG}|
\end{align*}

\subsection{\lpbase: The Basic Algorithm of \system}

\label{subsec:basic:algorithm}

\system takes as input a query $Q$ (Eq.~\eqref{eq:cq}) and a set of
statistics on the input database consisting of $\ell_p$-norms on
degree sequences, and returns: (1) a numerical upper
bound $B$ such that $|Q|\leq B$ whenever the input database satisfies
these statistics; (2) an explanation consisting of a q-inequality on
$|Q|$, which, for the particular numerical values of the
$\ell_p$-norms implies $|Q| \leq B$; and (3) a proof of the Shannon
inequality needed to prove the q-inequality.
For that, \system solves a Linear Program (LP) called \lpbase defined as follows:

\smallskip

\noindent {\bf The Real-valued Variables} are all unknowns
$h(U)\geq 0$, $\forall U \subseteq \vars(Q)$ ($2^n$ real-valued
variables).

\smallskip

\noindent {\bf The Objective} is to maximize $h(V_0)$, where $V_0$ is the set of    
the \groupby variables of the query in Eq.~\eqref{eq:cq}, under the
following two types of constraints.

\smallskip

\noindent {\bf The Statistics Constraints} are linear constraints of
the form in Eq.~\eqref{eq:h:p}, one for each $\ell_p$-norm of a degree sequence
that has been computed on the input database.

  \nop{
    \begin{align}
      \frac{1}{p}h(U) + h(V|U) \leq & \log\left(\lp{\degree_R(V|U)}_p\right)\label{eq:stats:constraint}
    \end{align}
}
\smallskip

\noindent {\bf The Shannon Constraints} are all basic Shannon inequalities,
as linear constraints (Eq.~\eqref{eq:shannon:monotonicity}
and~\eqref{eq:shannon:submodularity}).

\system uses the off-the-shelf solver HiGHS 1.7.2~\cite{HiGHS:2018} to solve both
\lpbase and its dual linear program.  The optimal solution of \lpbase
 consists of $2^n$ values $h^*(U)$, one for each set of query
variables $U$.  The optimal solution of the dual consists of
non-negative weights $w^*\geq 0$, one for every statistics constraint, and
non-negative weights $s^* \geq 0$, one for each basic Shannon inequality.
\system returns the following: (1) The bound $B \defeq 2^{h^*(V_0)}$,
(2) the q-inequality
$|Q| \leq \prod \left(\lp{\degree_R(V|U)}_p\right)^{w^*}$ where the
product ranges over all statistics constraints: this uses only the
weights associated to the Statistics Constraints, and (3) all basic
Shannon inequalities together with their weight $s^*$: these form the
required proof of the q-inequality.  We prove in the full paper:

\begin{theorem} For any input query $Q$, \system is correct:
\begin{enumerate}
    \item The quantity $B$ returned by \system is a tight upper bound on $|Q|$, meaning that $|Q|$ never exceeds $B$ if the input database satisfies the given statistics, and there exists an input database satisfying the given statistics on which
  $|Q|$ is as large as $B$ (up to a small query-dependent
  constant).
    \item The q-inequality returned by $\system$ holds in
  general. For the particular values of the statistics \\
  $\lp{\degree_R(V|U)}_p$, the inequality implies $|Q| \leq B$. 
    \item The basic Shannon inequalities multiplied with their associated
  weights $s^*$ form the proof of the q-inequality.
\end{enumerate}
\end{theorem}

Recall that the statistics only use simple degree sequences; without this assumption the tightness statement no
longer holds.

\begin{example} Consider the 3-way join $J_3$, shown
  in~\eqref{eq:j3}, and assume that, for each relation $R, S, T$
  and each attribute, \system has access to five precomputed $\ell_p$
  norms: $\ell_1, \ell_2, \ell_3, \ell_4, \ell_\infty$.  (Notice that
  $\ell_1$ is the same as the cardinality:
  $\lp{\degree_R(*|Y)}_1=|R|$.)  Then the optimal bound to $J_3$ is
  given by the following \lpbase linear program, with $2^4=16$ variables
  $h(\emptyset), h(X), h(Y), \ldots, h(XYZU)$:
  \begin{align*}
    \texttt{max}&\texttt{imize } h(XYZU) \texttt{ subject to} \\
              & h(XY) \leq \log |R| \\
              & \frac{1}{2}h(Y)+h(X|Y) \leq \log\lp{\degree_R(X|Y)}_2\\
              & \hspace{10mm} \ldots \text{same for all other $\ell_p$ norms}\\
              & h(X)+h(Y)\geq h(XY)+h(\emptyset)\\
              & h(XY)+h(YZ) \geq h(XYZ)+h(Y)\\
              & \hspace{10mm} \ldots \text{continue with all basic Shannon inequalities}
  \end{align*}
  A standard LP package returns both the optimal of this LP, $h^*(U)$,
  and the optimal of its dual, $w^*,s^*$.  The query's upper bound is
  $2^{h^*(XYZU)}$.  The q-inequality is
  $|Q| \leq |R|^{w_1^*} \cdot \lp{\degree_R(X|Y)}_2^{w_2^*} \cdots$
  where $w_1^*, w_2^*, \ldots$ are the dual variables associated with
  the statistical constraints.  Finally, the dual variables $s^*$
  associated to the basic Shannon inequalities provide the proof of
  the information-theoretic inequality needed to prove the
  q-inequality.
\end{example}

\section{Improving the Estimation Time}
\label{sec:algorithm}

%

\system needs to compute the upper bound in milliseconds in order to
be of use for query optimization.  To achieve this, we start by
applying two simple optimizations to the Basic Algorithm \lpbase in
Sec.~\ref{subsec:basic:algorithm}, which, recall, uses $2^n$ numerical
variables.  (1) for each atom $R_j(V_j)$ of the query, we consolidate
all variables $X_i$ that do not occur anywhere else into a single
variable, and (2) we retain only the \emph{Elemental Basic Shannon
  Inequalities}\footnote{Eq.~\eqref{eq:shannon:submodularity} is
  elemental if it is of the form $h(X_iW)+h(X_jW)\geq h(X_iX_jW)+h(W)$
  where $X_i,X_j$ are single variables and $W$ a set of
  variables;~\eqref{eq:shannon:monotonicity} is elemental if it is of
  the form $h(V)\geq h(V-\set{X_i})$ where $V=\set{X_1, \ldots, X_n}$
  is the set of all variables.}~\cite{Yeung:2008:ITN:1457455} in the
list of constraints, which are known to be complete.  Even with these
optimizations, \lpbase takes 100 ms already for queries with
$n \approx 10$ logical variables $X_1, \ldots, X_n$
(Fig.~\ref{fig:LpBound-estimation-time}).  We describe below two
improvements. \nop{that scale to larger queries.  To simplify the
  presentation, we assume that $Q$ (Eq.~\eqref{eq:cq}) is a full
  conjunctive query.}

\subsection{\lptdb: Berge-Acyclic Queries}

Our first algorithm works under two restrictions: the query needs to
be Berge-acyclic (Sec.~\ref{sec:background}), and all degree
constraints must be full and simple.  These restrictions are actually
quite generous: the JOBjoin, JOBlight, JOBrange, and STATS benchmarks used in Sec.~\ref{sec:experiments}
satisfy them.  Recall that $V= \set{X_1, \ldots, X_n}$ are the
variables of the query $Q$, and $R_1(V_1), \ldots, R_m(V_m)$ are the
atoms of $Q$.  For each variable $X_i$, let $a_i$ denote the number of atoms
that contain it.  We denote by $E_Q$ the following entropic
expression:
\begin{align}
  E_Q = & \sum^m_{j=1} h(V_j) - \sum^n_{i=1} (a_i-1) h(X_i) \label{eq:expr:q}
\end{align}
The linear program called \lptdb is the following:

\smallskip

\noindent {\bf The Real-valued Variables} are $h(X_1),\ldots,h(X_n)$ and
$h(V_1),\ldots,$ $h(V_m)$. Thus, instead of $2^n$ real-valued variables
$h(U)$, we only have one for each query variable $X_j$, and one for
each set $V_j$ corresponding to an atom $R_j(V_j)$, for a total of
$m+n$.

\smallskip

\noindent {\bf The Objective} is to maximize $E_Q$, under the
following constraints.

\smallskip

\noindent {\bf Statistics Constraints:} all constraints in
Eq.~\eqref{eq:h:p} are included.  This is possible
because the degree sequence is full and simple, and LHS can be written
as $\frac{1}{p}h(U)+h(V|U) = h(UV) - \frac{p-1}{p}h(U)$, where $U$ is
a single variable, and $UV$ is the set of variables of some
relation.

\smallskip

\noindent {\bf Additivity Constraints:} instead of all Shannon
inequalities, we have $1+|V_j|$ constraints for each atom $R_j(V_j)$
(for a total of $m+\sum_j|V_j|$ constraints):
\begin{align*}
  h(V_j) \leq & \sum_{i: X_i \in V_j} h(X_i) &\text{and}\ \ \ \ \ h(X_i) \leq  &h(V_j), \forall X_i \in V_j
\end{align*}

\begin{theorem} \label{th:lpbase:eq:lptdb}
  The optimal values of \lpbase and \lptdb are equal.
\end{theorem}

The proof uses techniques from information theory and is included in the supplementary material. 

\begin{example}
  We illustrate \lptdb on the 3-way join query $J_3$ in
  Eq.~\eqref{eq:j3}, and assume for simplicity that the only available
  statistics are the cardinalities (i.e., the $\ell_1$-norm of any
  full degree sequence): $|R|=|S|=|T|=M$, therefore, the AGM bound
  applies: $|J_3|\leq |R|\cdot |T| = M^2$.  The \lptdb is the
  following (where $m \defeq \log M$):
  \begin{align*}
    & \texttt{maximize } E_{J_3}\defeq h(XY)+h(YZ)+h(ZU)-h(Y)-h(Z)\\
    & \text{subject to} \\
    & \ h(XY) \leq m,\ h(YZ) \leq m,\ h(ZU) \leq m\\
    & \ h(X)\leq h(XY),\ h(Y)\leq h(XY),\ h(XY)\leq h(X)+h(Y)\ // \text{for } R(XY)\\
    & \hspace{44mm} \text{similarly for $S(YZ)$ \text{ and } $T(ZU)$}
  \end{align*}
  Notice that we only use 7 real-valued variables: we do not have
  real-valued variables for $h(XYZ)$ or $h(YU)$ etc.  One optimal
  solution is $h^*(X)=\cdots=h^*(U)=m/2$, $h^*(XY)=h^*(YZ)=h^*(ZU)=m$,
  and $E_{J_3}^*=2m$, implying $|J_3|\leq M^2$.  Notice that the
  additivity constraints are important in order to obtain a tight
  bound: if we dropped them, then the linear program admits the
  feasible solution $h^{**}(X)=\cdots=h^{**}(U)=0$,
  $h^{**}(XY)=h^{**}(YZ)=h^{**}(ZU)=m$, and $E_Q^{**}=3m$, leading to
  a weaker bound $|J_3|\leq M^3$.  Thus, the additivity constraints
  are unavoidable.  Statistics beyond cardinalities can easily be
  added, for example, an $\ell_4$ constraint on $\degree_S(Z|Y)$
  becomes
  $\frac{1}{4}h(Y)+h(Z|Y) = h(YZ) - \frac{3}{4}h(Y) \leq \log
  \lp{\degree_S(Z|Y)}_4$.
\end{example}

\lptdb can be adapted to Berge-acyclic queries with group-by as follows. Given a Berge-acyclic query $Q$ with group-by variables $V_0$, we can derive an equivalent Berge-acyclic query $Q'$ by removing from $Q$ the variables that are not in $V_0$ and are not join variables. The full Berge-acyclic query $Q''$, which is obtained from $Q'$ by promoting all variables in $Q'$ to group-by variables, has output size at least that of $Q'$. The quantity returned by \lptdb for $Q''$ is thus a valid upper bound on the size of $Q$.
\begin{example}
    Consider again the Berge-acyclic group-by query StarG in Sec.~\ref{sec:lpbound-groupby-queries}. We rewrite it into 
    \begin{align*}
        \text{StarG}''(X_1,X_2,Z) = R_1(X_1,Z)\wedge R_2(X_2,Z)\wedge S'(Z)
    \end{align*}
    \lptdb maximizes the quantity
    $E_{|\text{StarG}''|} = h(X_1Z) + h(X_2Z) + h(Z) - 2h(Z)$, under
    statistics and additivity constraints.  This yields a better bound
    than~\eqref{eq:sg:bound}, because the statistics constraints
    imply:
    \begin{align*}
      \frac{1}{3}  \log |\dom(S.Z)|  & + \log \lp{\degree_{R_1}(X_1|Z)}_3   + \log \lp{\degree_{R_2}(X_2|Z)}_3 \\
 \geq & \frac{1}{3}h(Z) + \left(\frac{1}{3}h(Z)+h(X_1|Z)\right)+\left(\frac{1}{3}h(Z) + h(X_2|Z)\right)=E_{|\text{StarG}''|}
    \end{align*}
    which leads to the following q-inequality, improving
    over~\eqref{eq:sg:bound}:
    \begin{align*}
      |\text{StarG}| \leq |\text{StarG}''| 
      \leq |\dom(S.Z)|^{1/3}\cdot \lp{\degree_{R_1}(X_1|Z)}_3\cdot\lp{\degree_{R_2}(X_2|Z)}_3
    \end{align*}
%
\end{example}


\nop{, which are norms on non-full degree constraints of the original relations.}

\subsection{\lpflow: Using Network Flow}
\label{subsec:lpflow}

\begin{figure}[t]
  \centering
  \begin{tikzpicture}[scale = 0.9, every node/.style={transform shape, scale = 1, inner sep = 1.2}]
      \node [draw, circle, double] (phi) at (0,0) {$\emptyset$};
      \node [draw, circle] (X) at (2,+2) {$X$};
      \node [draw, circle] (Y) at (2,0) {$Y$};
      \node [draw, circle] (Z) at (2,-2) {$Z$};
      \node [draw, circle] (XY) at (4,+1) {$XY$};
      \node [draw, circle] (YZ) at (4,-1) {$YZ$};
      \draw (phi) [->] to node[midway, sloped, above]{$w_1$} (X);
      \draw[->, bend left = 15] (X) to node[midway, sloped, above]{$w_2$} (XY);
      \draw[->, gray, densely dotted, bend left = 15] (XY) to node[midway, sloped, below]{$\infty$} (X);
      \draw[->, gray, densely dotted, bend left = 0] (XY) to node[midway, sloped, below]{$\infty$} (Y);
      \draw[->, bend left = 15] (Y) to node[midway, sloped, above]{$w_3${\color{red}${+w_4}$}} (YZ);
      \draw[->, gray, densely dotted, bend left = 15] (YZ) to node[midway, sloped, below]{$\infty$} (Y);
      \draw[->, gray, densely dotted, bend left = 0] (YZ) to node[midway, sloped, below]{$\infty$} (Z);
      \draw (phi) [->, very thick, red] to node[midway, sloped, below=-0.05]{${w_4/2}$} (Y);
    \end{tikzpicture}
  \caption{Example for \lpflow.}
  \label{fig:lpflow}
  \vspace*{-1em}
\end{figure}

Our second algorithm works for {\em any} conjunctive query (not
necessarily acyclic), and {\em any} constraints (they need to be on
\emph{simple}: recall that we only consider simple degree sequences in
this paper).  Our algorithm consists of a new linear program, \lpflow,
that uses a number of real-valued variables that is quadratic in the
query size: this is much better than the exponential number in
\lpbase, and slightly worse than the linear number in \lptdb. \lpflow
reduces the problem to a collection of network flow problems. It
generalizes the flow-based linear program introduced
in~\cite{DBLP:journals/corr/abs-2211-08381} for the \maxdegree bound
to the general statistics considered by \system.

%
\lpflow is different from both \lpbase and \lptdb. We
describe it only on an example, which illustrates both the original algorithm from~\cite{DBLP:journals/corr/abs-2211-08381}, and our generalization to $\ell_p$-norms. An in-depth account is given in the supplementary material.

\begin{example}
    \label{ex:lpflow}
    Consider the 2-way join $J_2$ in Eq.~\eqref{eq:j2}, along with statistics $|\dom(R.X)|$, $\lp{\degree_R(Y|X)}_\infty$, and
    $\lp{\degree_S(Z|Y)}_\infty$.  Our target is to find coefficients
    $w_1, w_2, w_3$ that make the following q-inequality valid and
    minimize the bound:
\begin{align}
    |J_2| \leq
        |\dom(R.X)|^{w_1} \cdot
        \lp{\degree_R(Y|X)}_\infty^{w_2} \cdot
        \lp{\degree_S(Z|Y)}_\infty^{w_3}
    \label{eq:lpflow:j2}
\end{align}
For that, the following needs to be a valid information inequality:
\begin{align}
    h(XYZ) \leq w_1 h(X) + w_2 h(Y|X) + w_3 h(Z|Y)
    \label{eq:lpflow:j2:ineq}
\end{align}
The key insight from~\cite{DBLP:journals/corr/abs-2211-08381} is that
checking the validity of such inequality (where all degree constraints
are {\em simple}) is equivalent to constructing a flow network
$G=(\nodes,\edges)$, and checking whether each variable $X, Y, Z$ is
independently receiving a maximum flow of at least 1.  In our
example, the flow network $G$ is shown in Fig.~\ref{fig:lpflow}
(ignore the \textbf{\color{red}red} part referring to ${\color{red}w_4}$ for now), where the nodes are the source $\emptyset$, individual variables $\{X\}$, $\{Y\}$, $\{Z\}$,
and sets $\{X, Y\}, \{Y, Z\}$ corresponding to available degree sequences
$\degree_R(Y|X), \degree_S(Z|Y)$.  The edges are of two types:
\begin{itemize}
    \item \textbf{Forward edges} like $X \to XY$ with capacity $w_2$. This represents the term $w_2 h(Y|X)$.
    \item \textbf{\color{gray}Backward edges} like $XY \to X$ with capacity $\infty$. This represents the monotonicity $h(X) \leq h(XY)$.
\end{itemize}

For inequality~\eqref{eq:lpflow:j2:ineq} to be valid,
\begin{itemize}
    \item $X$ needs to receive a flow of at least $1$.  Intuitively,
      this means $w_1 \geq 1$.  
    \item Independently, $Y$ needs to receive a flow of at least
      $1$. Intuitively, there is only one path from the source
      $\emptyset$ to $Y$, which is $\emptyset \to X \to XY \to Y$, and
      this implies that $\min(w_1, w_2, \infty) \geq 1$.  Formally,
      however, we need to setup a standard network flow LP: there is
      one flow variable \nop{for} $f_{a,b}$ for each edge $(a,b)$, with a
      capacity constraint $f_{a,b} \leq w_{a,b}$, and there is one
      flow-preservation constraint for each node (other than source
      and target); e.g., the constraint at node $X$ is
      $f_{\emptyset,X}+f_{XY,X}-f_{X,XY}=0$.
    \item Independently, $Z$ needs to receive a flow of at least
      $1$. Similar to above, this implies that
      $\min(w_1, w_2, \infty, w_3, \infty) \geq 1$, but formally we
      need a separate network flow LP.
\end{itemize}
To capture all network flows using a single LP, we simply create three
separate real-valued flow variables for each edge $(a, b)$, namely
$f_{a,b; X}, f_{a,b; Y}, f_{a,b; Z}$.  The \lpflow is shown below 
(ignore the {\bf\color{red} red} text referring to {\color{red}$w_4$} for now): 
  \begin{align}
    \min\quad
        &w_1 \log |\dom(R.X)|+
        w_2 \log \lp{\degree_R(Y|X)}_\infty\label{eq:lpflow:j2:lp:revised}\\
        &\quad +w_3 \log \lp{\degree_S(Z|Y)}_\infty
        {\color{red}+w_4 \log \lp{\degree_S(Z|Y)}_2}\nonumber\\
    \text{s.t.}\quad&w_1, w_2, w_3, {\color{red}{w_4}} \geq 0\nonumber\\
      & (f_{a,b;X})_{(a,b) \in \edges} \text{ form a flow $\emptyset\rightarrow X$ of capacity $\geq 1$}\nonumber\\
      & (f_{a,b;Y})_{(a,b) \in \edges} \text{ form a flow $\emptyset\rightarrow Y$ of capacity $\geq 1$}\nonumber\\
  \end{align}
  \begin{align}
        & (f_{a,b;Z})_{(a,b) \in \edges} \text{ form a flow $\emptyset\rightarrow Z$ of capacity $\geq 1$}\nonumber\\
        & f_{\emptyset,X;*}\leq w_1,\quad
        f_{X,XY;*}\leq w_2,\quad
        f_{Y,YZ;*}\leq w_3 {\color{red}+ w_4},\nonumber\\
        &{\color{red}f_{\emptyset,Y;*}\leq w_4/2}\nonumber
  \end{align}

There are $3n\sum_j |V_j|$ total variables, because the network has
$2\sum_j|V_j|$ edges, and for each edge $(a,b)$ we need to create one
capacity variable $w_{a,b}$, and $n$ real-valued variables:
$f_{a,b;X_i}$, $i=1,n$.

%
%
%

We next outline how to generalize the above algorithm to handle bounds
on arbitrary $\ell_p$-norms of degree sequences.  Continuing with the
above example, suppose that we are additionally given  $\lp{\degree_S(Z|Y)}_2$.  The RHS of the
q-inequality~\eqref{eq:lpflow:j2} now has an additional factor of
$\lp{\degree_S(Z|Y)}_2^{w_4}$ where $w_4$ is a new coefficient.
Similarly, the RHS of inequality~\eqref{eq:lpflow:j2:ineq} now has two
additional terms $+\frac{w_4}{2}h(Y) + w_4 h(Z|Y)$.  Accordingly, the
flow network from Fig.~\ref{fig:lpflow} is extended with extra
edges, depicted in \textbf{\color{red}red}.  In particular, we
have an extra edge from $\emptyset$ to $Y$ with capacity $w_4/2$, and
an extra edge from $Y$ to $YZ$ adding a capacity of $w_4$, on top of
the existing capacity of $w_3$.  These extra edges lead to new paths
that can be used to send flow to $Y$ and $Z$.  As a result, the
objective function of the above linear program is
extended with the red term.  The capacity constraints on the red
edges also change: They become $f_{\emptyset, Y; X} \leq w_4/2$,
$f_{\emptyset, Y; Y} \leq w_4/2$, $f_{\emptyset, Y; Z} \leq w_4/2$,
and similarly $f_{Y,YZ;X} \leq w_3+w_4$ etc.
\end{example}
The above bound can be straightforwardly generalized to handle \groupby by only considering flows $f_{a, b; X_i}$ where $X_i$ is a \groupby.

We prove the following theorem in the supplementary material.
\begin{theorem} \label{th:lpbase:eq:lpflow}
    The optimal values of \lpbase and \lpflow are equal.
  \end{theorem}

\subsection{Putting them Together}

Given a query $Q$, \system checks if $Q$ is Berge-acyclic and if all
statistics are full (they are always simple), and, in that case it
uses \lptdb to compute the bound, since its size is only linear in the
size of $Q$ and the statistics.  Otherwise, it uses \lpflow, whose
size is quadratic in the size of the query.  

\section{Support for Selection Predicates}
\label{sec:histograms}

\system can support arbitrary selection predicates on a relation. As long as we can provide $\ell_p$-norms on the degree sequences of the join columns for those tuples that satisfy the selection predicate, \system can use these norms in the statistics constraints. In the following, we discuss the case of equality and range predicates, and their conjunction and disjunction; \texttt{IN} and \texttt{LIKE} predicates can be accommodated using data structures like for \safebound~\cite{SafeBound:SIGMOD23}.

As data structures to support predicates, \system uses simple and effective adaptations of existing data structures in databases: Most Common Values (MCVs) and histograms. Yet instead of a count for each MCV or histogram bucket, \system keeps a set of $\ell_p$-norms on the degree sequences of the tuples for that MCV or histogram bucket. The simplicity and ubiquity of these data structures make \system easy to incorporate in database systems.

In the following, let a relation $R({\bf X},{\bf Y},A)$ with join attributes ${\bf X} = \{X_1,\ldots,X_n\}$, a predicate attribute $A$, and other attributes ${\bf Y}$.

\paragraph{Equality Predicate.} For each MCV $a$ of $A$, we compute $\ell_p$-norms for the full and simple degree sequences $\deg_R(*|X_i, A=a)$ for $i=1,n$. The number of MCVs can significantly affect the accuracy of \system (Fig.~\ref{fig:lpbound-MCVs}), as it does for \safebound and \psql. 

We also construct one degree sequence ${\bf d}_i$ for all non-MCVs of $A$ and each $i=1,n$. Let $r_i$ be the maximum number of $X_i$-values per non-MCV of $A$ and ${\bf d}_i$ be the degree sequence of the $r_i$ largest degrees of $X_i$-values. \nop{We can construct the degrees in the sequence by inspecting individually each non-MCV or aggregate over all non-MCVs.} We compute a set of $\ell_p$-norms of each degree sequence ${\bf d}_i$. An alternative, more expensive approach is to compute $\ell_p$-norms for each non-MCV and take their max for each $p$. 

To estimate for the equality predicate $A=v$, we use the $\ell_p$-norms for the degree sequences $\deg_R(*|X_i,$ $A=v)$ if $v$ is an MCV. Otherwise, we use the $\ell_p$-norms for the degree sequences ${\bf d}_i$. 

\paragraph{Range Predicate.} Range predicates are supported in \system using a hierarchy of histograms: Each layer is a histogram whose number of buckets is half the number of buckets of the histogram at the layer below. We ensure that the histogram at each layer covers the entire domain range of the attribute $A$. For each histogram bucket with boundaries $[s_i,e_i]$, we create $\ell_p$-norms on the full and simple degree sequences $\deg_R(*|X_i,$ $A\in [s_i,e_i])$.

To estimate for the range predicate $A\in [s,e]$, we find the smallest histogram bucket that contains the range $[s,e]$ from the predicate and then use the $\ell_p$-norms from that bucket.

\paragraph{Multiple Predicates.} In case of a conjunction of  predicates, we take as $\ell_p$-norm the minimum of the $\ell_p$-norms for the predicates, for each $p$. This is correct as the records must satisfy all predicates and in particular the most selective one. In case of a disjunction, we take as the $\ell_p$-norm the sum of the $\ell_p$-norms for the predicates, for each $p$. This {\em computed} quantity upper bounds the {\em desired} $\ell_p$-norm of the degree sequence for those tuples that satisfy the disjunction of the predicates, yet we cannot compute the latter norm unless we evaluate the predicates. To see this, observe that the desired $\ell_p$-norm is less than or equal to the $\ell_p$-norm of the degree sequence, which is obtained by the entry-wise sum of the degree sequences for the predicates. By Minkowski inequality, the latter norm is less than or equal to the computed norm.

\nop{\color{green}
In case of a disjunction, we take as $\ell_p$-norm the sum of the $\ell_p$-norms for the predicates, for each $p$. This is correct: the $\ell_p$-norm of the degree sequence for the disjunction of two predicates is less than or equal to the $\ell_p$-norm of the sum of two degree sequences, where the largest degrees of the two sequences are summed in order, which leads to the largest possible $\ell_p$-norm. Then, this $\ell_p$-norm of the sum of two degree sequences is less than or equal to the sum of the $\ell_p$-norms of the two degree sequences due to the Minkowski inequality, which proves that we take the upper bound.
}

\paragraph{Optimizations.} A challenge for \system is to estimate the cardinality of a join, where one operand is orders of magnitude larger than the other operands and has many dangling key values. This happens when a join operand has a selective predicate. By using norms that incorporate degrees  of dangling key values, \system returns a large overestimate. To address this challenge, it combines two orthogonal optimizations:  {\em predicate propagation} and {\em prefix degree sequences}. 
Predicate propagation is used in case of a predicate on a primary-key (PK) relation that is joined with a foreign-key (FK) relation. We propagate the predicate and its attribute through the join to the FK relation without increasing its size. The new predicate on the FK relation is then supported using MCVs and histograms to yield smaller and more accurate $\ell_p$-norms. \revtwo{For instance, assume we have a table $R(K,A)$ with primary key $K$ and attribute $A$ on which we have a predicate $\phi(A)$. We also have a table $S(K,B)$ with foreign key $K$ and some attribute $B$. By propagating $\phi(A)$ from $R$ to $S$, we mean that we join the two relations to obtain a new relation $S'(K,B,A)$. This relation $S'$ has the same cardinality as $S$, yet every $K$-value in $S'$ is now accompanied by the $A$-value from $R$. We can now construct MCVs and histograms on the data column $A$ in $S'$. A variant of this optimization is also used by \safebound~\cite{SafeBound:SIGMOD23}.
}

In case of a large degree sequence, \system also keeps its length ($\ell_0$-norm) and the $\ell_p$-norms on its prefixes with the $2^i$ largest degrees, for $i\geq 0$. Then, for a join, \system first fetches the $\ell_0$-norm of each of the operands. The minimum $m$ of these $\ell_0$-norms tells us the maximum number of key values that join at each operand. \system uses $m$ to pick the $\ell_p$-norms for the $i$-th prefix\footnote{The degrees typically decrease exponentially and sequence prefixes for $i>4$ have norms close to those for the entire degree sequence. For each large degree sequence, we therefore only keep the norms for the first 4 prefixes and for the entire sequence.} of the degree sequences of each of the join operands, for $2^{i-1}\leq m \leq 2^i$.

\nop{

TODO.  Things to say here:

\begin{itemize}
\item For a given attribute $Z$, we compute and store the maximum of
  the $\ell_p$ statistics of all relations of the form
  $\sigma_{Z=z}(R)$, for $z \in R.Z$.  We use these whenever the query
  contains a selection $R.Z=??$.  TO GIVE A NAME TO THIS STATISTICS,
  e.g. the \emph{generic conditionals}.
\item Better: construct a histogram on $R.Z$, by partitioning the
  domain into $b$ buckets, and storing separate generic conditionals
  per bucket.
\item Better: for the most common values $z_1, \ldots, z_k$ we store
  the values the $\ell_p$-statistics separately.  TO SAY HOW LARGE WE
  TAKE $k$.
\item For range queries, we need a different kind of histogram.
  Describe. 
\item For LIKE predicates we use this data structure XXX.
\item I really like Haoze's idea of the $\ell_p$ norms of a prefix.
  Can we include that?  Do we have experiments for that?
\item what else?
\end{itemize}

Relation $R(X,A_1,A_2,\ldots)$, conjunction of predicates $\bigwedge_i$ \textcolor{mMediumBrown}{$(A_i\text{ op } v_i)$}
\vspace*{1em}

Use the following $\ell_p$-norms for the conjunction of predicates:     \vspace*{1em}
\begin{itemize}
    \item Fetch the $\ell_p$-norms for each predicate \textcolor{mMediumBrown}{$A_i\text{ op } v_i$}
    \vspace*{1em}
    \item For each $p$, take the minimum of the $\ell_p$-norms for the predicates
\end{itemize}

}

\section{Experimental Evaluation}
\label{sec:experiments}

In this section, we experimentally answer the following questions: {\em How accurate are \system's cardinality estimates? Are \system's estimation time and space requirements sufficiently low for it to be practical? Can \system's  estimates  help avoid inefficient query plans, in case faster query plans exist?}
Our findings are as follows. 

1. \system can be orders of magnitude more accurate than traditional estimators used in mainstream open-source and commercial database systems. Yet \system has low estimation time (within a couple of ms) and space requirements (a few MBs), which are comparable with those of traditional estimators.

2. Learned estimators can be more accurate than \system, according to the errors reported in a prior extensive benchmarking effort~\cite{CE:VLDB21}\footnote{These estimation models are copyrighted and not available. Training and tuning the models requires knowledge that is not available (confirmed by authors of \cite{CE:VLDB21}).}. This is by design as their models are trained to overfit the specific dataset and possibly the query pattern. Downsides are reported in the literature, including: poor generalization to new datasets and query patterns; non-trivially large training times (including hyper-parameter tuning) and extra space, even an order of magnitude larger than the dataset itself~\cite{CE:VLDB21}.

3. By configuring \psql to use the estimates of \system for the 20 longest-running queries in our benchmarks, we obtained faster query plans than those originally picked by \psql.

\subsection{Experimental Setup}

\paragraph{Competitors.} 
We use the {\em traditional estimators} from open-source systems \psql 13.14 and \duckdb 0.10.1 and a commercial system \dbx. We use two {\em pessimistic cardinality estimators}: \safebound \cite{SafeBound:SIGMOD23} and our approach \system. \metarev{We use two classes of {\em learned cardinality estimators}: (1) The {\em PGM-based cardinality estimators} \bayescard~\cite{bayescard}, \deepdb~\cite{deepdb}, and \factorjoin~\cite{FactorJoin:SIGMOD23}; and (2) the {\em ML-based estimators} \flatcard~\cite{flat} and \neurocard~\cite{neurocard}. For the latter, we refer to their performance as reported in~\cite{CE:VLDB21}. We checked with 
the authors of \safebound, \bayescard, and \factorjoin that we used the best configurations for their systems and for \deepdb.}

\paragraph{Benchmarks.} Table~\ref{tab:queries} shows the characteristics of the queries used in the experiments. They are based on the benchmarks: JOB~\cite{DBLP:journals/vldb/LeisRGMBKN18} over the IMDB dataset (3.7GB); STATS\footnote{\url{https://relational-data.org/dataset/STATS}} over the Stats Stack Exchange network dataset (38MB); and SM (subgraph matching) over the DBLP dataset (26.8MB edge relation and 3.5MB vertex relation)~\cite{SubgraphMatching:SIGMOD20}. The SM queries are cyclic, all other queries are acyclic.
For IMDB, we use JOBlight and JOBrange queries from previous work~\cite{DBLP:conf/cidr/KipfKRLBK19, neurocard}, which have both equality and range predicates. We further created JOBjoin queries without predicates. We also created JOBlight-gby, JOBrange-gby and STATS-gby queries, which are JOBlight, JOBrange and STATS queries with group-by clauses consisting of at most one attribute per relation\footnote{\metarev{This is not a restriction, \system can support arbitrary group-by clauses. This is our methodology for generating query workloads with GROUP-BY.}}: We classify them into three groups of roughly equal size: small domain (domain sizes of the group-by attributes are $\leq 150$); large domain (domain sizes $>$ 150); and a mixture of both. The SM queries use 11-28 copies of the edge relation and 2 vertex relation copies per edge relation copy, with one equality predicate per vertex relation copy.

\begin{table}[t]
    \centering
\hspace*{-1em}\begin{tabular}{|l|r|r|r|l|}\hline
Benchmark & \#queries & \#rels & \#preds & query type \\\hline
JOBjoin & 31 & 5-14 & 0 & snowflake \& full\\
JOBlight& 70 & 2-5  & 1-4 & star \& full \\
JOBrange & 1000 & 2-5 & 1-4 & star \& full \\
JOBlight-gby & 170 & 2-5 & 1-4 & star \& group-by \\
JOBrange-gby & 877 & 2-5 & 1-4 & star \& group-by \\\hline
STATS  & 146 & 2-8 & 2-16 & acyclic \& full \\
STATS-gby & 370 & 2-8 & 2-16 & acyclic \& group-by \\\hline
SM     & 400 & 33-84 & 22-56 & cyclic \& full \\\hline
\end{tabular}
    \caption{Benchmarks used in the experiments.}
    \label{tab:queries}
    \vspace*{-1em}
\end{table}

\nop{
We considered three classes of full acyclic JOB queries: JOBjoin (31 snowflake queries over 5-14 relations, no predicates); JOBlight (70 star queries over 3-5 relations, with 1-4 equality predicates); and JOBlightranges (1000 star queries over 3-5 relations, with 1-4 range predicates). We considered all 146 full acyclic STATS queries over 2-8 relations with 2-16 predicates. 
We also created new queries with group-by clauses consisting of at most one attribute per relation. We classify them into three groups: small domain (the group-by attributes have domain sizes $\leq 150$); large domains (domain size $>$ 150); and a mixture of both. Overall, we created $67 + 70 + 70 = 207$ and $124 + 121 + 125 = 370$ group-by queries in  JOBLight and respectively STATS.
SM has 400 full cyclic queries joining 11-28 copies of the edge relation and two vertex relations per copy of the edge relation, with one equality selection predicate per vertex relation.
}

\paragraph{Metrics.} We report the estimation error, which is the estimated cardinality divided by the true cardinality of the query output. The estimation error is greater (less) than one in case of over (under)-estimation. We report the (wall-clock) estimation time of the estimators. We also report the end-to-end query execution time of the 20 longest-running queries using \psql when injected the estimates of some of the estimators. We also report the extra space needed for the data statistics and ML models used for estimation.

\paragraph{System configuration.} We used an Intel Xeon Silver 4214 (48 cores) with 193GB memory, running Debian GNU/Linux 10 (buster). For \psql, we used the recommended configuration~\cite{DBLP:journals/vldb/LeisRGMBKN18}: 4GB shared memory, 2GB work memory, 32GB implicit OS cache, and 6 max parallel workers. We enabled indices on primary/foreign keys. We used the default configuration for data statistics for each estimator. \system uses HiGHS 1.7.2~\cite{HiGHS:2018} for solving LPs.

\begin{figure}[t]
    \centering
    \includegraphics[width=.75\textwidth]{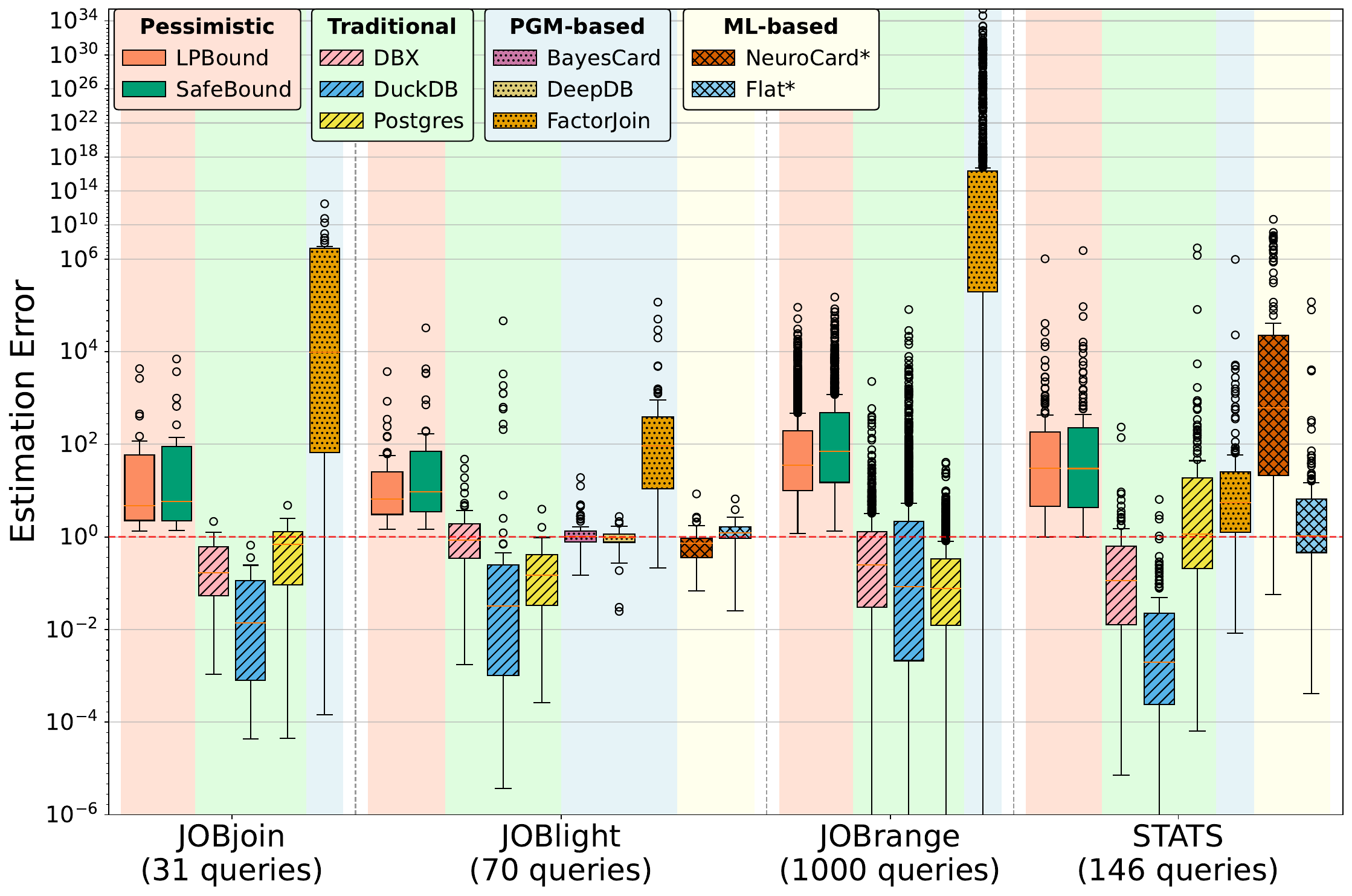}
    \caption{Estimation errors for JOBJoin, JOBLight, JOBRange, and STATS. For the starred ML-based estimators, we use the errors for JOBLight and STATS reported in the literature~\cite{CE:VLDB21}.
    }
    \label{fig:estimates-combined}
\end{figure}

\begin{figure*}[h!]
\centering
\includegraphics[width=.95\textwidth]{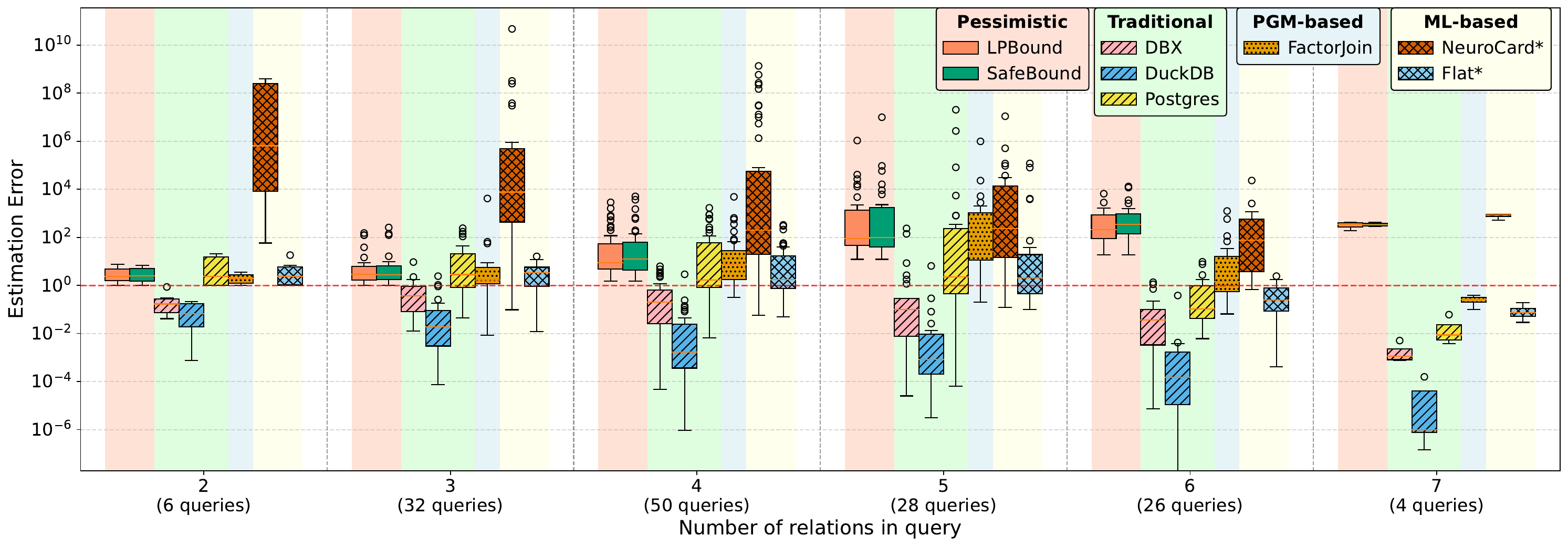}
\caption{Estimation errors for STATS. For the starred ML-based estimators, we use errors reported in the literature~\cite{CE:VLDB21}.}
\label{fig:estimates-STATS}
\end{figure*}

\begin{figure}[t]
    \centering
    \begin{minipage}{0.49\textwidth}
        \centering
        \includegraphics[width=\textwidth]{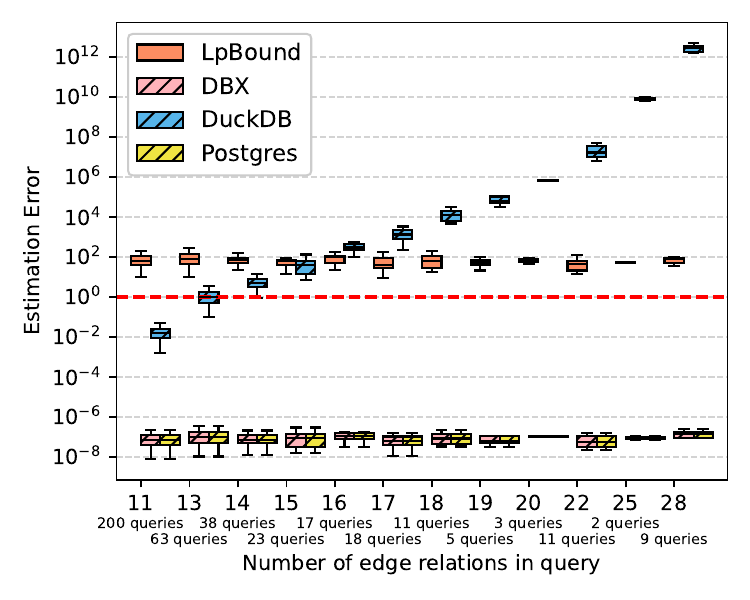}
        \vspace*{-1cm}
        \caption{Estimation errors for the SM cyclic queries.}
        \label{fig:cyclic-estimation-error}
    \end{minipage}
    \hfill
    \begin{minipage}{0.49\textwidth}
        \centering
        \includegraphics[width=\textwidth]{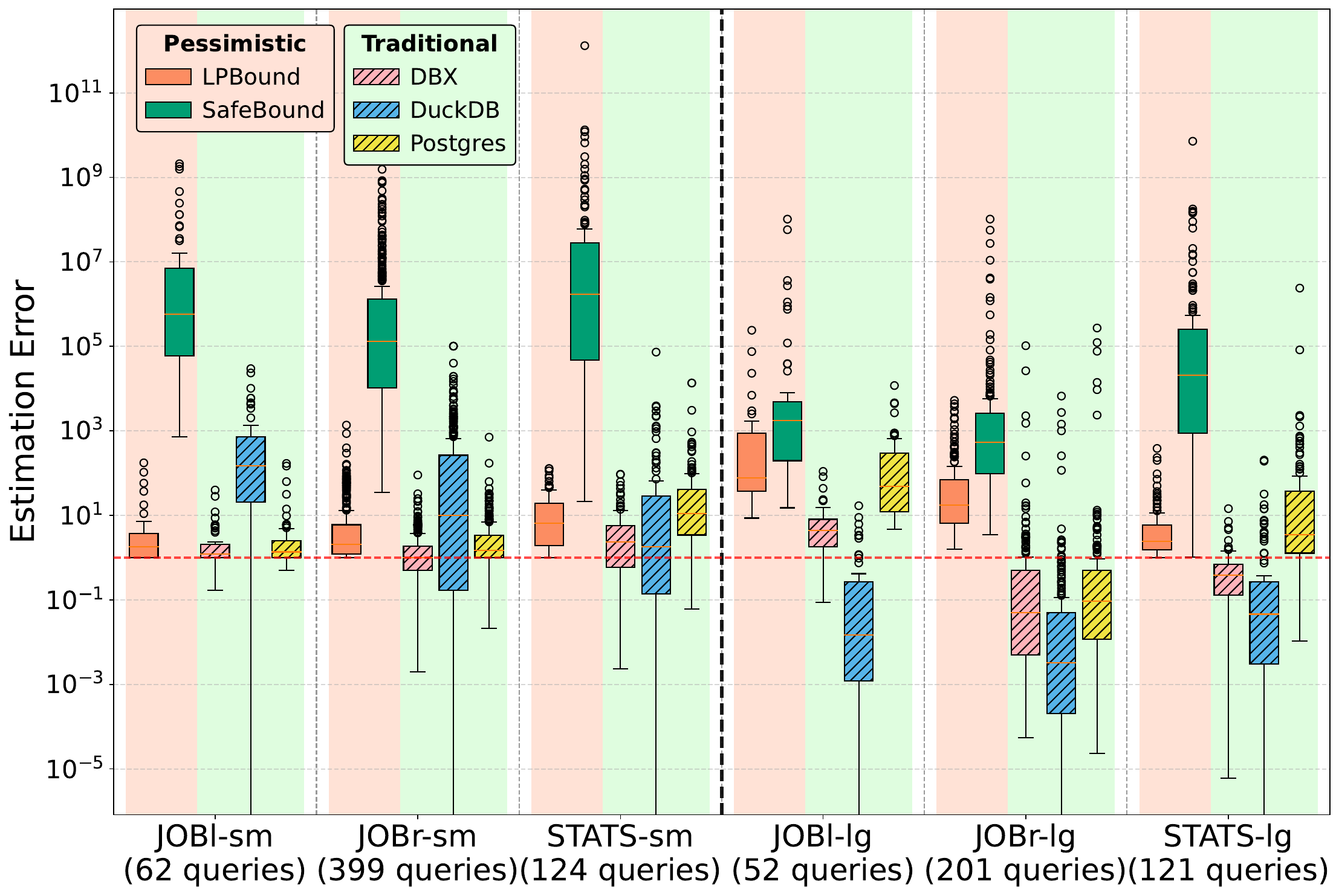}
        \vspace*{0.2cm}
        \caption{Estimation errors for group-by queries.}
        \label{fig:groupby-estimation-error}
    \end{minipage}
\end{figure}

\begin{figure*}[t]
    \centering
    \begin{minipage}[b]{0.48\textwidth}
        \centering
        \includegraphics[width=\textwidth]{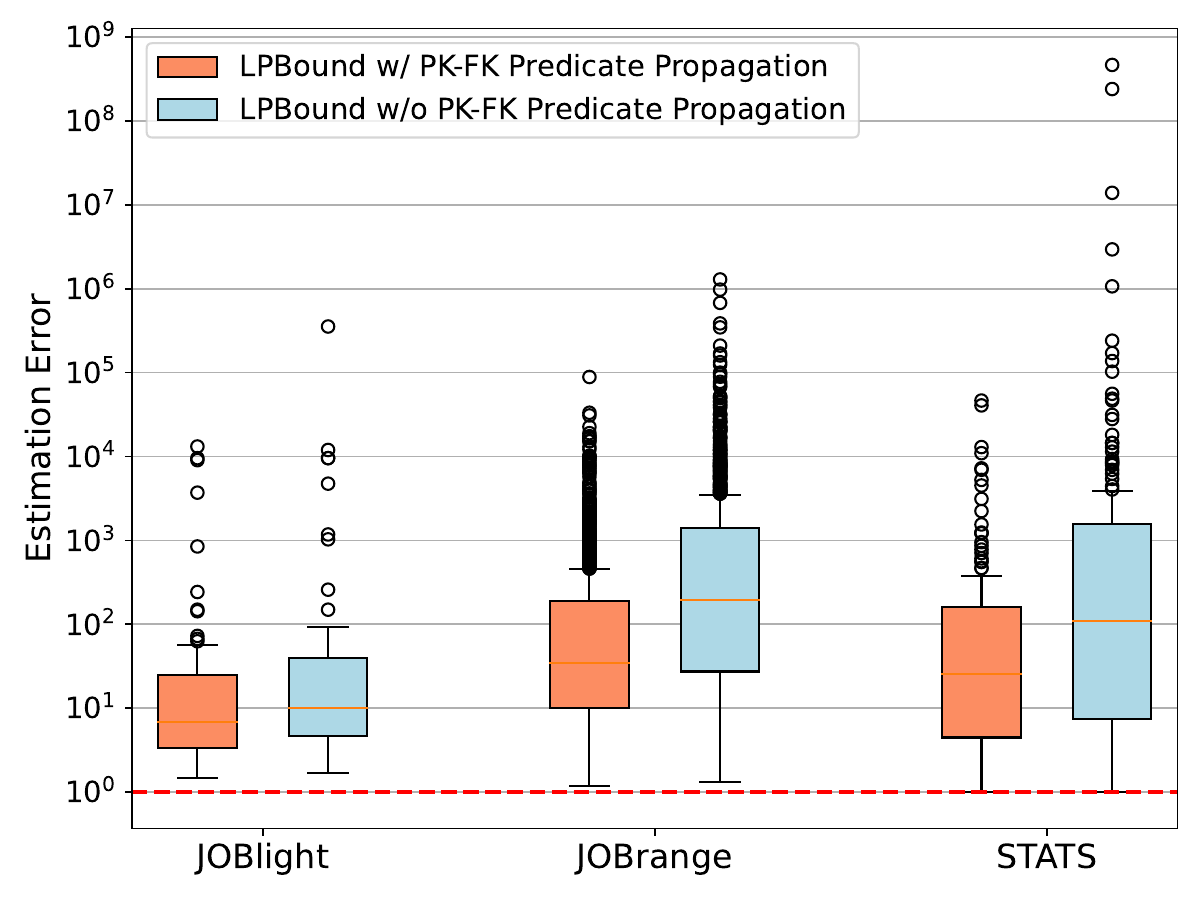}
    \end{minipage}
    \hfill
    \begin{minipage}[b]{0.48\textwidth}
        \centering
        \includegraphics[width=\textwidth]{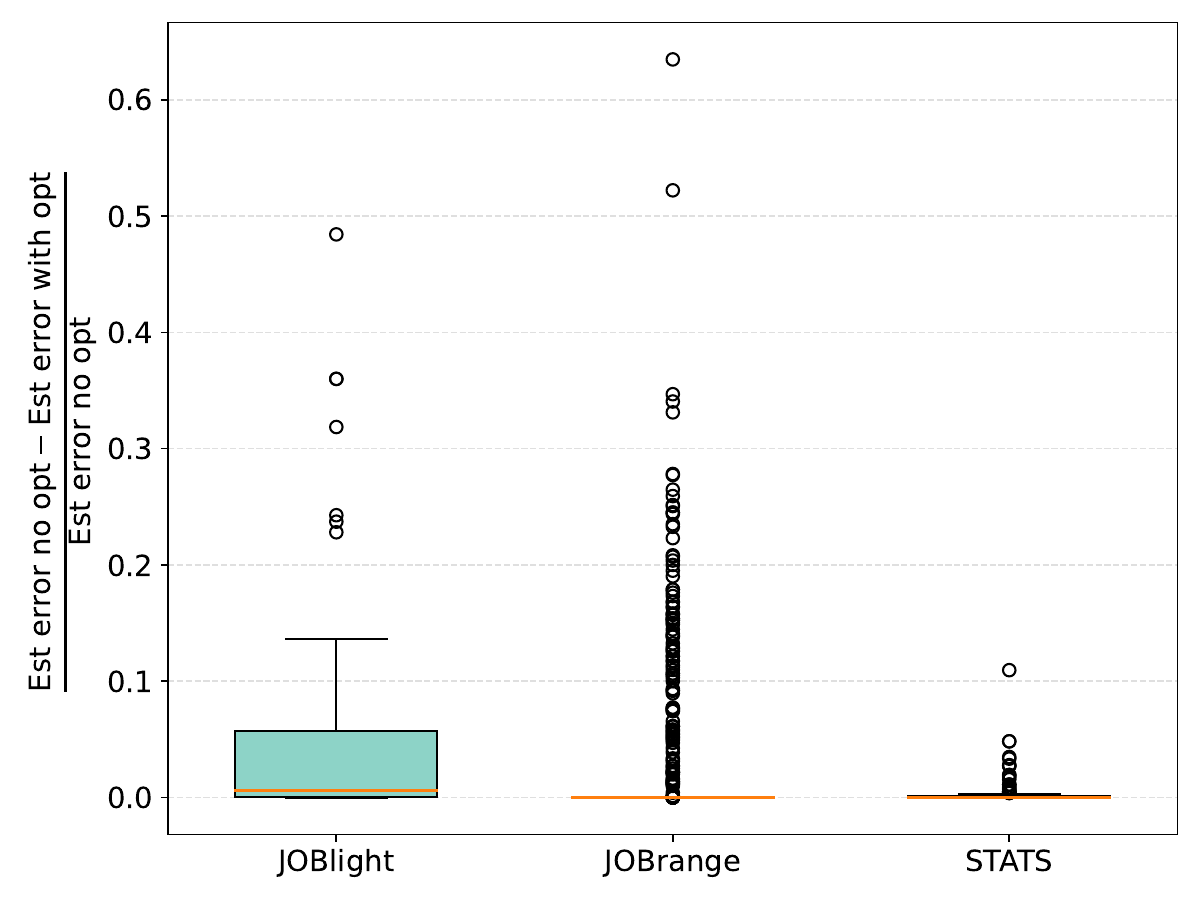}
    \end{minipage}
    \caption{Improvements on estimation errors when using the two optimizations discussed in Sec.~\ref{sec:histograms}. Left: PK-FK predicate propagation optimization. Right: Prefix degree sequences optimization.}
    \label{fig:improvements-optimizations}
\end{figure*}

\subsection{Estimation Errors}

\paragraph{Acyclic queries.} {\em \system has a smaller error range than the traditional estimators and \safebound for acyclic queries.} Fig.~\ref{fig:estimates-combined} plots the estimation errors for the acyclic queries. All systems except \system and \safebound both underestimate and overestimate. The traditional estimators broadly use as estimation the multiplication of the relation sizes and of the selectivities of the query predicates. The selectivity of a join predicate is the inverse of the minimum of the domain sizes of the two join attributes (so average degree, as opposed to maximum degree, is used). For equality and range predicates, Most Common Values (for \psql) and histograms (for \psql and \dbx) are used. \duckdb has a fixed selectivity of 0.2 for a range predicate. For ML-based estimators, we use the estimates reported in~\cite{CE:VLDB21}, as the models are not available. These models were designed to overfit JOBlight and subsequently fine-tuned to STATS, albeit with a poorer accuracy. 

We also report on the estimation errors of the PGM-based estimators. \bayescard and \deepdb do very well on JOBlight; this is the only workload on which their implementation works. \factorjoin builds high-dimensional probability distributions over the attributes of each relation to capture their correlation. This building task uses random sampling for JOB and the more accurate \bayescard for STATS.
\factorjoin faces a trade-off between good accuracy and fast estimation time. To keep the latter practical, it approximates the learned high-dimensional distributions by the product of one-dimensional distributions for JOB\footnote{The implementation of \factorjoin  does not support 2D distributions for JOB.} and of two-dimensional distributions for STATS. These choices influence the estimation error: It is far more accurate for STATS than for JOBjoin due to the choice of \bayescard over sampling and 2-dimensional over 1-dimensional factorization. The errors for JOBrange are larger possibly due to the larger number (up to 3) of predicates per relation.

Fig.~\ref{fig:estimates-STATS} shows that the accuracy of the estimators decreases with the number of relations per query (shown for STATS, a similar trend also holds for JOBlight and JOBrange): The traditional estimators underestimate more, whereas the pessimistic estimators overestimate more. \neurocard starts with a large overestimation for a join of two relations and decreases its estimation as we increase the number of relations; the other ML-based estimators follow this trend but at a smaller scale.

\paragraph{Cyclic queries.} {\em \system is the most accurate estimator for the SM cyclic queries.} Fig.~\ref{fig:cyclic-estimation-error} shows the errors of \system and the traditional estimators, grouped by the number of edge relations in the query. The learned estimators do not work for cyclic queries.\footnote{The implementation of \factorjoin does not support SM queries.}

As we increase linearly the number $n$ of edge relations from 11 to 28, the number of join conditions between the edge relations increases quadratically in $n$. This poses difficulties to the traditional estimators, which exhibit two distinct behaviors. 

{\em The estimate of \psql and \dbx is 1 for all SM queries and their error is the inverse of the query output size. This is an underestimation by 7-8 orders of magnitude.} The estimation is obtained by multiplying: the size of the edge relation $n$ times; the selectivity of each of the $n^2$ join conditions; the size of the vertex relation $2n$ times; the selectivity of the $2n$ join conditions between the edge and vertex relations; and the selectivity of the $2n$ equality predicates in the $2n$ vertex relations. The product of the relation sizes is much smaller than the inverse of the product of these selectivities, so the estimation is a number below 1, which is then rounded to 1.

{\em The estimate of \duckdb increases exponentially in the number of edge relations, eventually leading to an overestimation by over 12 orders of magnitude.} 
Its estimation ignores most of the join conditions, but accounts for each of the $n$ copies of the edge relation, as explained next. To estimate, \duckdb first constructs a graph, where each node is a relation in the query and there are two edges between any two nodes representing relations that are joined in the query: one edge per attribute participating in the join. Each edge is weighted by the inverse of the domain size of the attribute. \duckdb then takes a minimum-weight spanning tree of this graph. A significant factor in the  estimation is then the multiplication of the ($n$ edge and $2n$ vertex) relation sizes at the nodes and of the weights of the edges in the spanning tree ($3n-1$ domain sizes of one or the other column in the edge or vertex relations). For each of the relations in the query, the estimate has thus a factor proportional to the fraction of the relation size over an attribute's domain size.

\paragraph{Group-by queries.} {\em The range of the estimation errors for group-by queries is the smallest for \system.}
Except for \system, \psql, and \dbx, the systems ignore the group-by clause and estimate the cardinality for the full query. 
Fig.~\ref{fig:groupby-estimation-error} shows the errors for the small and large domain classes of JOBlight, JOBrange, and STATS group-by queries (the mixed domain class behaves very similarly to the large domain class). For small domain sizes (first half of figure), \system and \psql use the product of the domain sizes, which is close to the true cardinalities. For large domain sizes (second half), the true cardinalities remain smaller than for the full queries, yet \psql estimates are for the full queries. This explains why the error boxes are shifted up relative to those in Fig.~\ref{fig:estimates-combined}. \safebound and \duckdb estimate the full query and have large errors.

\paragraph{Optimization Improvements.}
Fig.~\ref{fig:improvements-optimizations} shows the improvements to the estimation accuracy brought by each of the two optimizations discussed in Sec.~\ref{sec:histograms}, when taken in isolation.

The left figure shows that,  when propagating predicates from the primary-key relation to the foreign-key relations, the estimation error can improve by over an order of magnitude in the worst case (corresponding to the upper dots in the plot) and by roughly 5x in the median case (corresponding to the red line in the boxplots). 

The right figure shows that,  when using prefix degree sequences for the degree sequences of relations without predicates, the estimation error can improve by up to 50\% for JOBlight queries, up to 65\% for JOBrange queries and up to 10\% for STATS queries. The improvement is measured as the division of (i) the difference between the estimation error without this optimization and the estimation error with this optimization and (2) the the estimation error without this optimization.

\begin{table}[t]
\centering
\small
\setlength{\tabcolsep}{2.5pt}
\renewcommand{\arraystretch}{1.1}
\begin{tabular}{|l|r|r|r|r|r|r|}
\hline
\multirow{2}{*}{\textbf{Estimator}} & \multicolumn{2}{c|}{\textbf{JOBjoin}} & \multicolumn{2}{c|}{\textbf{JOBlight}} & \multicolumn{2}{c|}{\textbf{STATS}} \\
\cline{2-7}
& Time & Space & Time & Space & Time & Space \\
\hline
\system & 0.48 / 10.5 & 0.04 & 0.36 / 1.5 & 1.25 & 0.49 / 1.6 & 3.62 \\
\safebound & 0.85 / 147.9 & 0.07 & 1.28 / 13.0 & 1.75 & 1.89 / 5.6 & 5.94 \\
\hline
\dbx$^{\!+}$ & - / 371.7 & - & - / 35.3 & - & - / 13.3 & - \\
\duckdb$^{\!+}$ & - / 99.4 & - & - / 535.2 & - & - / 30.3 & - \\
\psql$^{\!+}$ & - / 19.8 & $<$0.001 & - / 3.4 & 0.001 & - / 18.7 & 0.011 \\
\hline
\factorjoin & 0.66 / 202 & 31.6 & 16.7 / 166.5 & 22.8 & 35.3 / 626 & 8.2 \\
\metarev{\bayescard} & - / - & - & \metarev{3.0 / 21.7} & \metarev{1.6} & \nop{5.8 / -} - / - & \nop{5.9} - \\
\metarev{\deepdb} & - / - & - & \metarev{4.3 / 28.6} & \metarev{34.0} & \nop{87.0 / -} - / - & \nop{162.0} - \\
\hline
\neurocard$^{\!*}$ & - / - & - & 18.0 / - & 6.9 & 23.0 / - & 337.0 \\
\flatcard$^{\!*}$ & - / - & - & 8.6 / - & 3.4 & 175.0 / - & 310.0 \\
\hline
\end{tabular}
\caption{
Time (ms): average wall-clock times to compute estimates for (i) a sub-query of a query, averaged over all sub-queries of queries / (ii) a query and all its connected sub-queries, averaged over all queries.
Space (MB): extra space for data statistics and models. 
The times ($^+$) are for the entire query optimization task.
The numbers ($^*$) are from prior work~\cite{CE:VLDB21} and only available for JOBlight and STATS. (-) means unavailable data or unsupported workload.
}
\label{tab:time-space}
\end{table}
\subsection{Estimation Times}

{\em \system has a very low estimation time (a few ms) thanks to its LP optimizations. At the other extreme, ML-based estimators can be 1-2 orders of magnitude slower even when taking their average estimation time per subquery instead of the estimation time for all subqueries.}

To produce a plan for a query with $n$ relations, a query optimizer  uses the cardinality estimates for some of the $k$-relation sub-queries for $2\leq k \leq n$. Following prior work~\cite{CE:VLDB21,FactorJoin:SIGMOD23}, we use the sub-queries  produced by \psql's planner for a given query. The range (min-max) of the number of sub-queries is: 8-2018 for JOBjoin; 1-26 for JOBlight; and 1-75 for STATS. The times for JOBrange are not reported, but we expect them to be close to those for JOBlight. SM is not supported by the ML-based estimators and \safebound.

We report two estimation times per benchmark: (i) the time to compute the estimate for a single sub-query, averaged over all sub-queries of all queries, and (ii) the time to compute the estimates for a query and all its sub-queries, averaged over all queries.

\nop{The type (i) time represents the expected time to compute the estimate for an individual (sub-)query, while the type (ii) time represents the expected time needed for the cardinality estimation in the query optimization task for a query.}

Table~\ref{tab:time-space} reports the estimation times of the estimators. It was not possible to get the estimation times for the traditional estimators, so we report instead their times for the entire query optimization task to give a context for the other reported times; thy should have the lowest estimation times. The type (i) times for the starred ML-based estimators are from~\cite{CE:VLDB21}. We expect their type (ii) times to be at least an order of magnitude larger than their type (i) times, given the average number of sub-queries per query.

\factorjoin computes the estimates for all sub-queries of a query in a bottom-up traversal of a left-deep query plan. This computation does not parallelize well, however. Its type (i) time is therefore much lower than the type (ii) time. 

\system can effectively parallelize the LP solving for the sub-queries of a query. Even though one can extend an already constructed LP to accommodate new relations and statistics, we found that it is faster to avoid estimation dependencies between the related sub-queries and estimate for them independently in parallel. \nop{\system naturally exploits the observation that smaller LPs are solved faster than larger LPs, so it distributes the workload accordingly to the available threads.}

\nop{
Table~\ref{tab:time-space} reports the times needed by the \system, \safebound, and \factorjoin estimators to compute the estimates for the entire space of investigated sub-queries of a query, averaged over all queries in a benchmark. 
It was not possible to get such estimation times for the traditional estimators, so we report instead their average times for the entire query optimization task to give a context for the other times in the table; their estimation times alone are expected to be the lowest among all competitors. 
The times for the ML-based estimators, except \factorjoin, are the averages over all sub-queries of each query~\cite{CE:VLDB21}. We expect their overall estimation times to be at least an order of magnitude larger, given the average number of sub-queries per query.
The times for \factorjoin are computed by us. \factorjoin can naturally share the estimate computation for all subqueries, so its overall time is very close to the time to estimate for the full query.
}

\subsection{Space Requirements}
\label{sec:experiments-space-requirements}

{\em The extra space used by \system for statistics is 1.6x less than of \safebound and 1.2-93x less than of the ML-based estimators.}

Table~\ref{tab:time-space} shows the amount of extra space needed to store the data statistics or machine learning models used by the  estimators. 

The traditional estimators use modest extra space.
\psql uses 100 MCVs per predicate column: Increasing the number of MCVs leads to very large estimation time, as it computes the join output size at estimation time for the MCVs. It also uses 100 buckets per histogram and sampling-based estimates of domain sizes for columns. \duckdb only uses (very accurate and computed using hyperloglog~\cite{hyperloglog,DBLP:conf/edbt/HeuleNH13}) domain size estimates, no MCVs, and no histograms. \dbx uses histograms with 200 buckets and no MCVs. 

\nop{Running \texttt{analyze} improves significantly the domain size estimates in \psql.}

\safebound uses a compressed representation of the degree sequences and 2056 MCVs on the predicate columns. 

\system uses up to\footnote{Only 2/8 predicate attributes have domain sizes (134k, 235k) greater than 2k in JOBlight; for JOBrange, there are 3/13 such domains (15k, 23k, 134k). For STATS, the domain sizes are at most 100. For SM, the predicates are on the label attribute from the vertex relation and with domain size 15. Each edge relation joins with two copies of the vertex relation, so we use two predicates to indirectly filter the edge relation. We use $15\times 15$ MCVs to capture all possible combinations of the two predicates.} 5000 MCVs on the predicate columns in JOB, albeit for less space than \safebound. Both \safebound and \system use hierarchical histograms on data columns with 128 buckets. \system stores $\ell_p$-norms within each histogram bucket, while \safebound stores $\ell_1$-norms (counts) only. Although not reported in the table, \system needs 1.12MB for SM and 8MB for JOBrange. JOBrange has queries with more predicates, which need support, and more columns with large domains.

The models used by \neurocard and \flatcard are a feature-rich representation of the datasets. They take more space than the statistics used by the other estimators. For STATS, these models take 10x more space than the dataset itself~\cite{CE:VLDB21}.

\begin{table}[t]
    \centering
\small
\setlength{\tabcolsep}{2.5pt}
\renewcommand{\arraystretch}{1.1}
\metarev{
\begin{tabular}{|l|r|r|r|r|r|}
\hline
  {\bf Estimator}   & {\bf JOBjoin} & {\bf JOBlight} & {\bf JOBrange} & {\bf STATS} & {\bf SM}\\\hline
   \system-${\ell_1}$     & 4.07 & 12.35 & 42.42 & 24.67 & 0.65 \\
   \system     & 14.95 & 19.06 & 54.79 & 27.91 & 1.59 \\
   \safebound  & 88.56 & 162.09  & 209.23  & 32.04  &  - \\\hline
   \factorjoin & 10068.8 &  4990.9 &  5042.7 & 360.92 & - \\
   \bayescard  & -    & 493.36 & -    & - & - \\
   \deepdb     & -    & 1191.17   & - & - & - \\\hline
   \neurocard$^*$  & -    & 3600 & - & - & - \\ 
   \flatcard$^*$  & -     & 3060 & - & - & - \\\hline
\end{tabular}
}
    \caption{Time (sec) to compute the required statistics for the pessimistic and PGM-based estimators. 
    \system-${\ell_1}$ is \system with $\ell_1$-norms only.
    (-) means the system cannot estimate for the respective workload. (*) means the times  are from prior work~\cite{CE:VLDB21}, as the code is not available.}
    \label{tab:times-compute-statistics}
\end{table}

\metarev{
\subsection{Time to Compute the Statistics}
\label{sec:experiments-stats-compute-time}

Table~\ref{tab:times-compute-statistics} gives the times to compute the statistics or models required by the estimators. 
Overall, the compute time for \system is at least one order of magnitude smaller\footnote{All estimators in Table~\ref{tab:times-compute-statistics} except \system compute their statistics using Python.} than for the PGM-based estimators. 
The computation of the statistics used by \system is fully expressed in SQL and executed using \duckdb. Such statistics are: the MCVs, the histograms, the $\ell_p$-norms ($p\in\{1,\ldots,10,\infty\}$) for each MCV, histogram bucket, and full relation, and the two optimizations (FKPK and prefix) from Sec.~\ref{sec:histograms}. About 80\% of \system's time is spent on the two optimizations. To better understand the effect of the number of norms, we also report the times for \system when restricted to the $\ell_1$-norm only. This shows that increasing from 1 to 11 norms only increases the compute time 1.5--3.7 times. 
\deepdb and \bayescard take 86\% and respectively 67\% of their compute time for training, the remaining time is for constructing an auxiliary data structure to support efficient sampling.  \factorjoin spends most of its time (98\%) to construct statistics to speed up the estimation, while relatively very short time (2\%) is spent on training the model. The times for \neurocard and \flatcard are as 
reported in prior work~\cite{CE:VLDB21} for training without hyper-parameter tuning.
}

\subsection{From Cardinality Estimates to Query Plans}

\begin{figure*}[t]
    \centering
    \includegraphics[width=.95\textwidth]{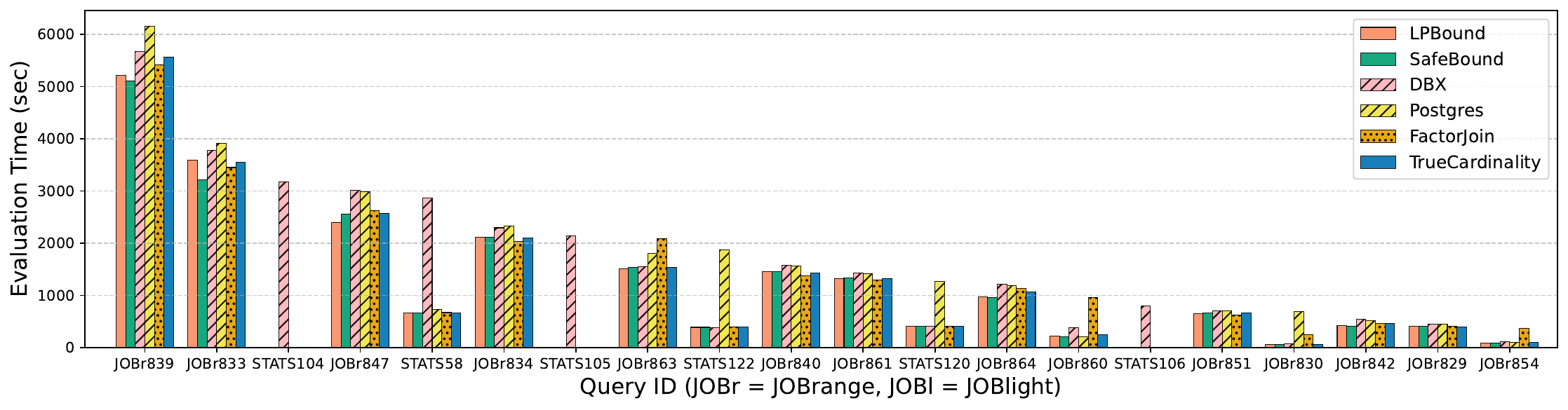}
    \caption{\psql (wall-clock) evaluation time for the 20 most expensive queries in JOBlight, JOBrange, and STATS,  when injected the estimates of \system, \safebound, \dbx, and \psql or the trues cardinalities for all subqueries of the query.  The runtimes for STATS 104, 105, 106 are very small when using the estimates of all systems but \dbx and therefore not visible.}
    \label{fig:most-expensive-queries}
\end{figure*}

\begin{figure*}[t]
    \centering
    \begin{minipage}[b]{0.48\textwidth}
        \centering
        \includegraphics[width=\textwidth]{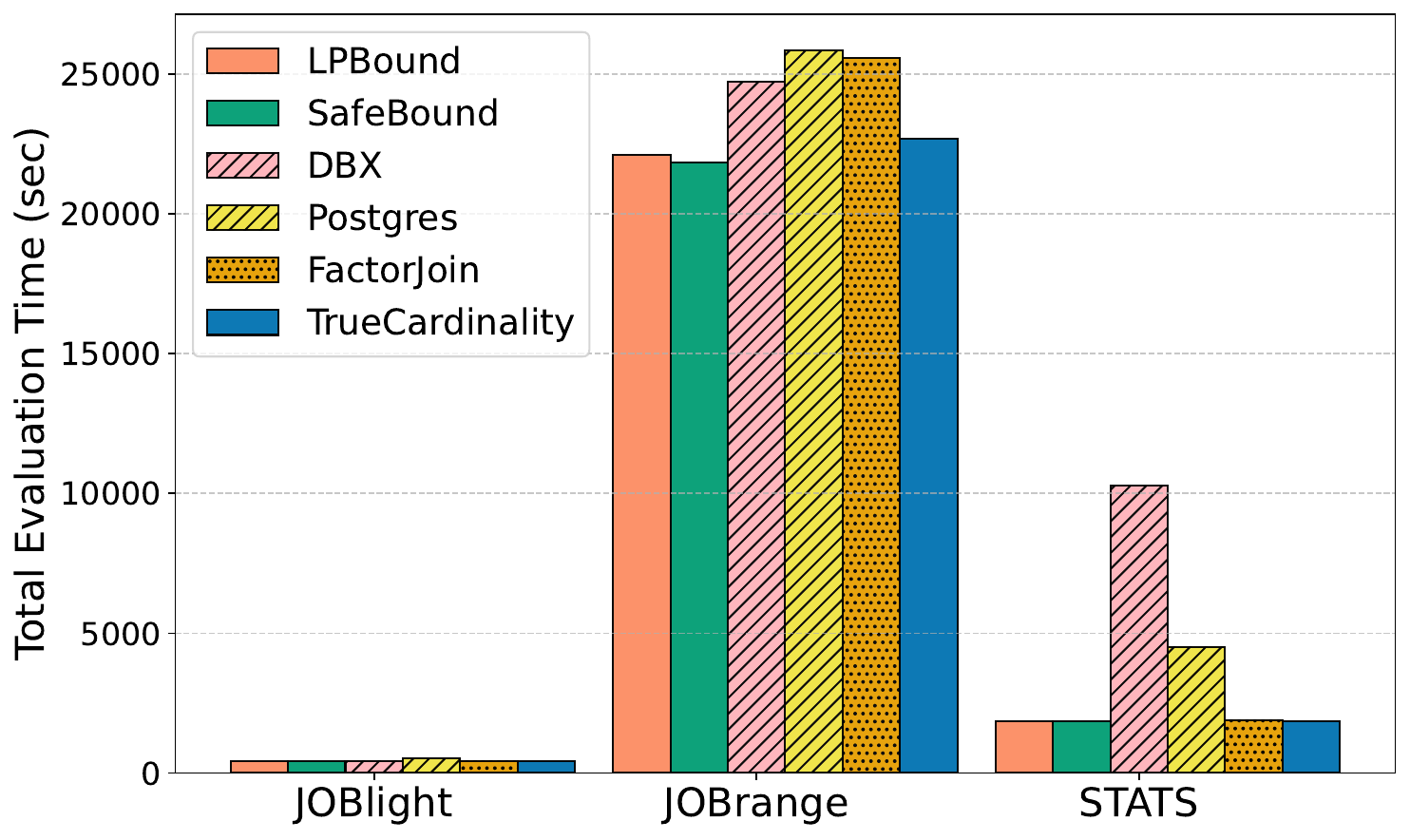}
    \end{minipage}
    \hfill
    \begin{minipage}[b]{0.48\textwidth}
        \centering
        \includegraphics[width=\textwidth]{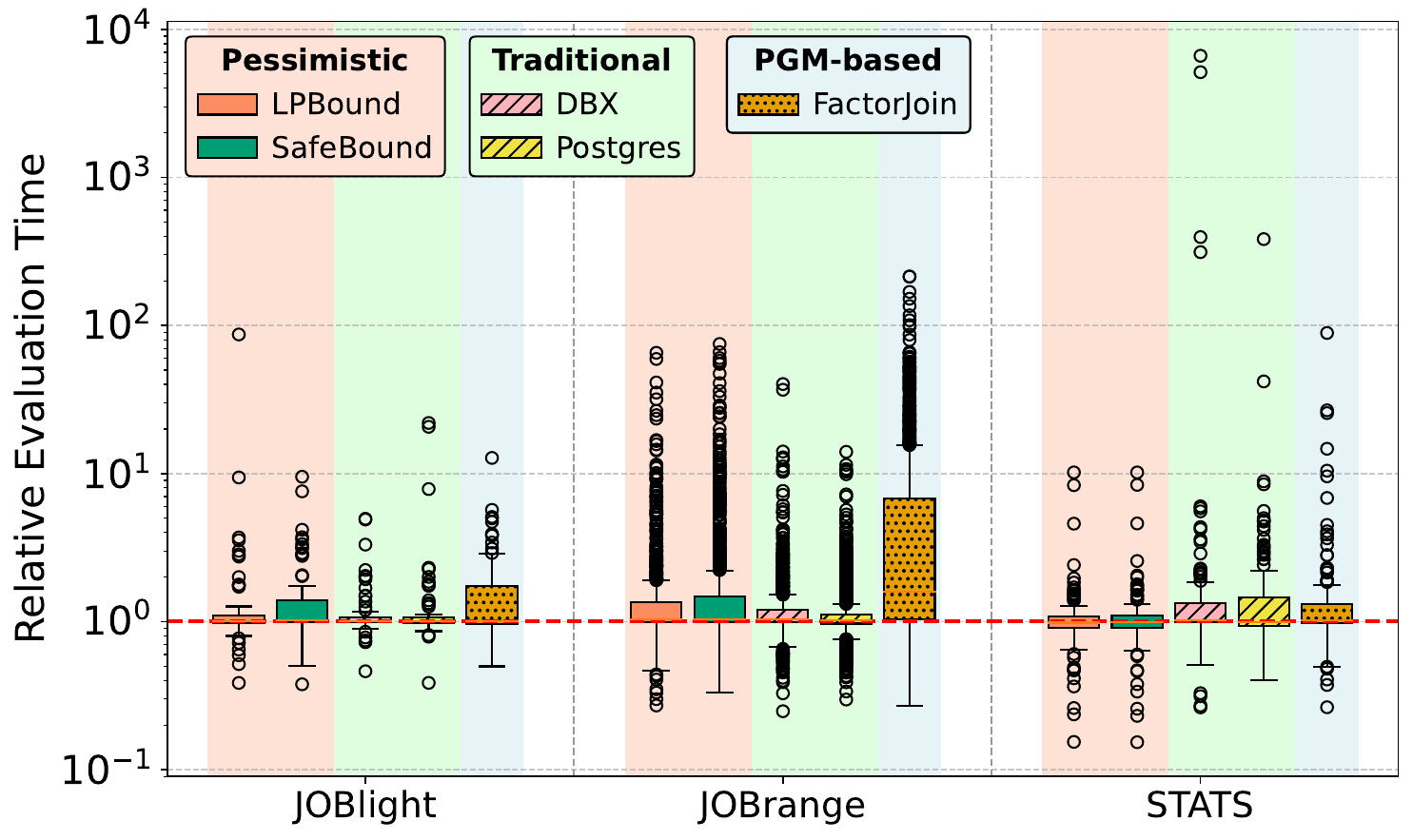}
    \end{minipage}
    \caption{Left: Overall evaluation time of all queries in a benchmark for \psql when using estimates for all subqueries from \system, \safebound, \dbx, \psql and true cardinalities. Right: Relative evaluation times compared to the baseline evaluation time obtained when using true cardinalities.}
    \label{fig:runtime}
\end{figure*}

{\em When injected the estimates of \system, \psql derives query plans at least as good as those derived using the true cardinalities.}

This result was expected and aligns with observations from prior work~\cite{DBLP:conf/sigmod/CaiBS19,SafeBound:SIGMOD23}.
We verified this for the 20 queries in JOBlight, JOBranges, and STATS (Fig.~\ref{fig:most-expensive-queries}), 
which took longest to execute using \psql when the query plan was generated based on the estimations of \system, \safebound, \dbx, \psql, or \factorjoin.
We used \psql for query execution as it easily allows to inject external estimates into its query optimizer\footnote{\url{https://github.com/ossc-db/pg_hint_plan}}. Remarkably, the estimates of \system can lead to better \psql query plans than using true cardinalities, e.g., for the 9 most expensive JOBrange queries in the figure. Prior work~\cite{DBLP:journals/vldb/LeisRGMBKN18} also reported this surprising behavior that using true cardinalities, \psql does not necessarily pick better query plans.
\safebound leads to better plans than \system for the top-2 most expensive queries. The estimates of \dbx and \psql lead in many cases to much slower query plans: for STATS104 (STATS122), \dbx (\psql) estimates lead to a plan that is more than 3000x (4x) slower than for the other estimators. For three STATS queries (104, 105, 106), the \dbx estimates yield a very slow plan; the runtimes of the plans using the estimates of the other systems are not visible in the plot.

Fig.~\ref{fig:runtime} (left) shows the aggregated \psql evaluation time of all queries in JOBlight, JOBrange, and STATS when using estimates for all sub-queries from \system, \safebound, \dbx, \psql, and true cardinalities (left).
Fig.~\ref{fig:runtime} (right)  shows the relative evaluation times compared to the baseline evaluation time obtained when using true cardinalities. 
We have two observations. First, overestimation can be beneficial for performance of expensive queries, which has been discussed in Section~\ref{sec:experiments}. Second, overestimation can be detrimental for performance of less expensive queries in some cases.

The first observation is reflected in the overall evaluation times, which are dominated by the most expensive queries in the benchmark (some of which are listed in Fig.~\ref{fig:most-expensive-queries}). 
Traditional approaches lead to higher evaluation times for the expensive queries, and therefore to higher overall evaluation times, while the pessimistic approaches lead to lower evaluation times for those expensive queries. Overall, the  evaluation times for the pessimistic approaches are about the same (JOBlight and STATS) or lower (JOBrange) than the baseline evaluation times.
The second observation is reflected in the relative evaluation times for the JOB benchmarks. 
The boxplots for the traditional approaches are lower than those for the pessimistic approaches, indicating that the traditional approaches perform better for the less expensive queries in the benchmarks.

\factorjoin has both high overall evaluation time and high relative evaluation time.
It estimates very accurately for the queries in STATS, thus has similar evaluation time to the baseline evaluation time. For the queries in JOBlight and JOBrange, it mostly overestimates, which leads to lower evaluation times for the expensive queries. However, the overestimations are significant, which makes it perform worse than the pessimistic approaches for the less expensive queries, as shown in the right plot of Fig.~\ref{fig:runtime}. This leads to the high overall evaluation time of \factorjoin.

\subsection{Performance Considerations for \system}
\label{ex:performance-considerations}

\begin{figure}[t]
    \centering
    \begin{minipage}{0.48\textwidth}
        \centering
        \includegraphics[width=\textwidth]{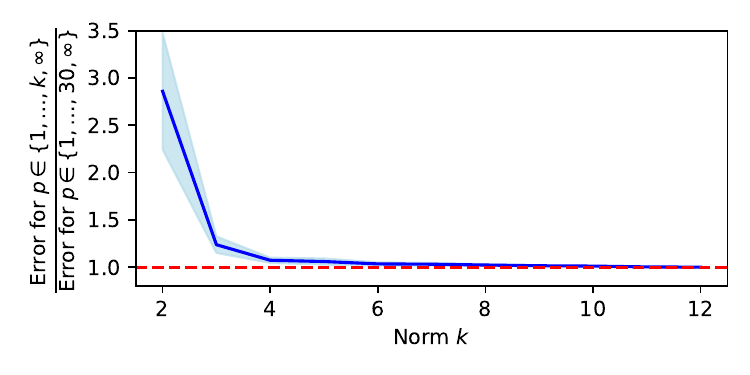}
        \caption{The amount of useful norms follows the law of diminishing returns: Plotting the division of estimation errors for the norms $\{1,\ldots,k,\infty\}$ and $\{1,\ldots,30,\infty\}$, averaged over the 70 JOBlight queries.}
        \label{fig:LpBound-amount-norms}
        \vspace*{-1em}
    \end{minipage}
    \hfill
    \begin{minipage}{0.48\textwidth}
    \centering
    \includegraphics[width=\textwidth]{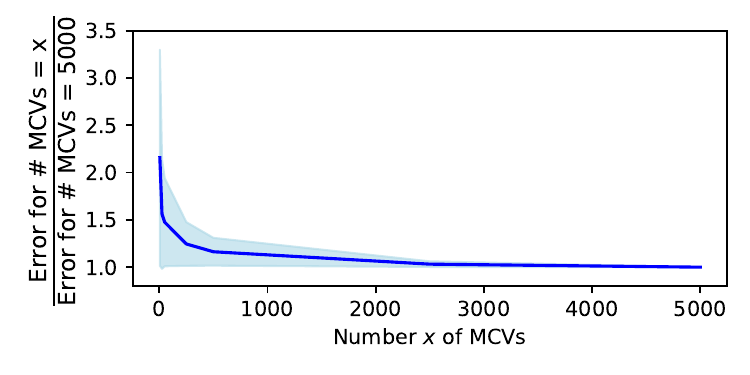}
    \caption{Effect of the number of MCVs on estimation error for \system on JOBlight.}
    \label{fig:lpbound-MCVs}
    \vspace*{-1em}
    \end{minipage}
\end{figure}

\paragraph{How Many $\ell_p$-Norms to keep?} Using the norms for $p\in[1,10]\cup\{\infty\}$ gives the best trade-off between the space requirements, the estimation error, and the estimation time for the JOBlight queries (Fig.~\ref{fig:LpBound-amount-norms}). This was verified to hold also for the other benchmarks. Further norms can still lower the estimation error, but only margina\-lly, and at the expense of more space and estimation time.

\paragraph{How many Most Common Values (MCVs)?} Using sufficiently many MCVs to support the estimation for selection predicates can effectively reduce the overall estimation error. For each of the top-$k$ MCVs of a predicate attribute, \system stores one set of $\ell_p$-norms. It also stores one further set of norms for all remaining attribute values (Sec.~\ref{sec:histograms}). 
\revthree{
For small-domain attributes, e.g., \texttt{COMPANY\_TYPE}, it is often feasible to have MCVs for each domain value. This significantly improves the estimation accuracy.
For large-domain attributes, e.g., \texttt{COMPANY\_ID}, it not not practical to do so. To decide on the number $k$ of MCVs, one can plot the estimation error as a function of $k$ and pick $k$ so that the improvement in estimation error for larger $k$ is below a threshold, e.g., $1\%$.
Fig.~\ref{fig:lpbound-MCVs} shows that  $k\leq 2500$ can yield on average to clear accuracy improvements for JOBlight; this is similar for JOBrange (not shown).}

\begin{figure}[t]
    \centering
    \includegraphics[width=.65\textwidth]{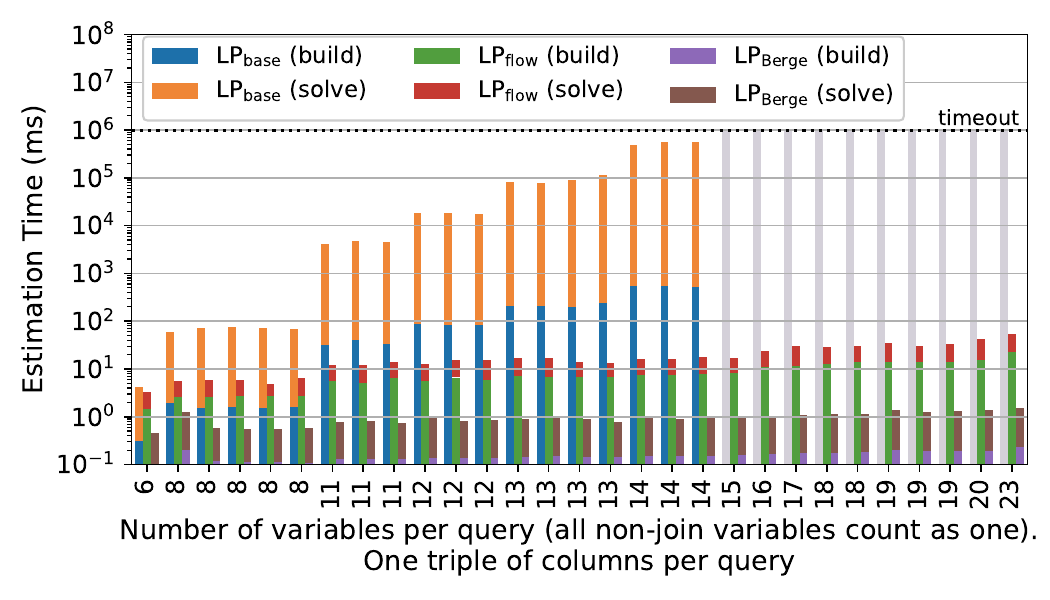}
    \caption{Estimation times for \system on full queries of JOBjoin using \lpbase and its optimizations \lpflow and \lptdb.}
    \label{fig:LpBound-estimation-time}
    \vspace*{-1em}
\end{figure}

\paragraph{Optimizations for \system's LPs.} The optimizations introduced for solving \system's LPs are essential for the practicality of \system. Fig.~\ref{fig:LpBound-estimation-time} shows the estimation time of \system using \lpbase, \lpflow, and \lptdb. We used JOBjoin, as its queries are Berge-acyclic and have the largest number of relations and variables among the considered benchmarks, and therefore can stress test and compare the efficiency of the three approaches. As expected, \lpbase takes too long (over 1000 seconds) to build and solve an LP with $2^{15}$ entropic terms and times out beyond this. \lpflow uses a network flow of size at most $15^2$ and finishes in under 70 ms for each JOBjoin query. Most of its time is spent constructing the network. \lptdb consistently takes under 2ms for all queries.

\section{Conclusion and Future Work}
\label{sec:conclusions}

In this paper we introduced \system, a pessimistic cardinality estimator that uses $\ell_p$-norms of degree sequences of the join columns and information inequalities. \revone{The advantages of \system over the learned estimators such as \factorjoin, \bayescard, and \deepdb, \neurocard,  and \flatcard  are that it provides: strong, one-sided theoretical guarantees; low estimation time and error when applied to workloads not seen before; fast construction of the necessary statistics; and a rich query language support with (cyclic and acyclic) equality joins, equality and range predicates, and group-by variables. This language support goes significantly beyond the star or even acyclic queries supported by the competing estimators benchmarked in Section~\ref{sec:experiments}.} 

\revthree{While \system's estimation time is slightly larger than that of  traditional estimators, it can nevertheless have lower estimation errors and lead to significantly improved query performance: Fig.~\ref{fig:most-expensive-queries} shows that the runtime improvement can be up to 3000 seconds for some queries, while its estimation time is only a few milliseconds.}

\revone{There are two major limitations of \system, as introduced in this paper. First, it can non-trivially overestimate the cardinality of joins of highly miscalibrated relations. We introduced two optimizations in Sec.~\ref{sec:histograms} to mitigate this problem. Second, it does not yet support range (and theta) joins, complex and negated predicates, and nested queries. }\metarev{\system's flexible framework can in principle accommodate such query constructs, yet this is not immediate and deserves an in-depth treatment in future work.
}
\metarev{
For example, an arbitrary predicate could be accommodated using appropriate data structures that can identify the ranges of tuples that satisfy the  predicate and that can be adjusted to store norms on the degree sequences within such ranges. 
A LIKE predicate can be accommodated, for instance, using a 3-gram index to select ranges of tuples that satisfy the predicate, similar to SafeBound~\cite{SafeBound:SIGMOD23}. To guarantee that \system returns an upper bound on the true cardinality of the query, the returned ranges must include all matching tuples.

To support nested queries, \system needs to become compositional, i.e., to take $\ell_p$-norms on input relations and return upper bounds on $\ell_p$-norms on the query output. Given a nested query $Q$, \system needs to first compute upper bounds on $\ell_p$-norms on the relations representing the sub-queries of $Q$ and then use these bounds to estimate the cardinality of $Q$.

Future work also needs to address the efficient maintenance of  \system's estimation under data updates. The following three observations outline a practical approach to achieve this. 
First, the q-inequality $|Q| \leq \prod \left(\lp{\degree_R(V|U)}_p\right)^{w^*}$, where the product ranges over all available statistics constraints, holds with {\em the same weights $w^*$} even when the norms $\lp{\degree_R(V|U)}_p$ change. This means that we do not need to solve the LP after every data update to obtain a valid upper bound on the cardinality of a query $Q$. 
Second, the $\ell_p$-norms used by \system can be expressed in SQL (as for Experiment~\ref{sec:experiments-stats-compute-time}) and maintained efficiently under data updates using the view maintenance mechanism of the underlying database system, e.g.,  delta queries. 
Third, we envision the use of $\ell_p$-sketches~\cite{cormode2020small} for an efficient, albeit approximate, maintenance of the $\ell_p$-norms.
}

\nop{
Our long term goal is to develop a framework for two-side approximations on the true cardinality of a query output, i.e., both guaranteed lower and upper bounds.
}

\bibliographystyle{plain} 
\bibliography{bib,bib-outliers}

\newpage

\onecolumn

\appendix

\section{Supplementary Material}

This is extra material for the submission titled
"\system: Pessimistic Cardinality Estimation Using
  $\ell_p$-Norms of Degree Sequences."
This material is organized as follows.
Section~\ref{subsec:proof:th:lpbase:eq:lptdb} gives the proof for Theorem~\ref{th:lpbase:eq:lptdb}.
Section~\ref{subsec:lptd} describes  a third optimized algorithm for estimating arbitrary conjunctive queries, which uses hypertree decompositions of the queries. This is not yet implemented in \system.
Section~\ref{subsec:lpflow:details} gives more details on the \lpflow optimization and the proof for  Theorem~\ref{th:lpbase:eq:lpflow}.
\nop{Section~\ref{app:predicates-examples} gives examples on how \system accommodates selection predicates.
Section~\ref{app:further-experiments} details further experiments.
}

\subsection{Proof of Theorem~\ref{th:lpbase:eq:lptdb}}
\label{subsec:proof:th:lpbase:eq:lptdb}
For simplicity of presentation, we assume here that the
query $Q$ is connected.
Denote by $b_{\text{base}}$ and $b_{\text{Berge}}$ the values of
\lpbase and \lptdb.  The inequality
$b_{\text{base}}\leq b_{\text{Berge}}$ follows from two observations:
\begin{itemize}
\item For any acyclic query $Q$ and any polymatroid $h$, the
  inequality $E_Q \geq h(X_1\cdots X_n)$ is a Shannon inequality.
  This is a well known
  inequality~\cite{DBLP:journals/tse/Lee87,DBLP:conf/sigmod/KenigMPSS20}
  (which we review in Lemma~\ref{lemma:tony:lee} below).
\item Any feasible solution to \lpbase can be converted to a feasible
  solution of \lptdb by simply ``forgetting'' the terms $h(U)$ that do
  not occur in \lptdb.
\end{itemize}

For the converse, $b_{\text{Berge}}\leq b_{\text{base}}$, we will
prove that every feasible solution $h$ to \lptdb can be extended to a
feasible solution to \lpbase (by defining $h(U)$ for all terms $h(U)$
that did not appear in \lptdb), such that $E_Q = h(X_1\cdots X_n)$.
We will actually prove something stronger: that $h$ can be extended to
a \emph{normal polymatroid}.

\begin{definition} \label{def:normal}
  A set function $h : 2^{\set{X_1,\ldots,X_n}} \rightarrow \R$ is a \emph{normal
    polymatroid} if $h(\emptyset)=0$ and it satisfies:
  \begin{align}
    \forall U \subseteq \set{X_1,\ldots,X_n}: \sum_{W\subseteq U}(-1)^{|W|+1}h(W) \geq &0 \label{eq:normal}
  \end{align}
\end{definition}
It is known that every normal polymatroid is an entropic vector, and
every entropic vector is a polymatroid, but none of the converse
holds.

The inequality $b_{\text{Berge}}\leq b_{\text{base}}$ follows from two lemmas:

\begin{lemma} \label{lemma:normal:extension:of:one:bag} Let
  $V = \set{X_1, \ldots, X_n}$, let $a_1, \ldots, a_n, A$ be $n+1$
  non-negative numbers such that:
  \begin{align}
    a_1 + \cdots + a_n \geq & A \ \ \ \ \text{and}\ \ \ \ \ a_i \leq A, \forall i=1,n \label{eq:a:a}
  \end{align}
  Then there exists a normal polymatroid $h : 2^V \rightarrow \R$ such
  that $h(X_i)=a_i$ for all $i=1,n$ and $h(V) = A$.
\end{lemma}

\begin{lemma}[Stitching Lemma] \label{lemma:stich} Let
  $V_1, V_2$ be two sets of variables, $Z \defeq V_1 \cap V_2$. Let
  $h_1:2^{V_1} \rightarrow \R$, $h_2: 2^{V_2} \rightarrow \R$ be two
  normal polymatroids that agree on their common variables $Z$: in
  other words there exists $h : 2^Z \rightarrow \R$ such that
  $\forall U \subseteq Z$, $h_1(U)=h(U)=h_2(U)$.  Define the following
  function $h' : 2^{V_1\cup V_2} \rightarrow \R$:
  \begin{align}
    h'(U) \defeq & h_1(U \cap V_1|U \cap Z) + h_2(U \cap V_2|U \cap Z) + h(U \cap Z) \label{eq:stich}
  \end{align}
  Then $h'$ is a normal polymatroid that agrees with $h_1$ on $V_1$ and
  with $h_2$ on $V_2$, and, furthermore, satisfies:
  \begin{align}
    h'(V_1 \cup V_2) = & h'(V_1) + h'(V_2) - h'(V_1\cap V_2) \label{eq:independent}
  \end{align}
  In essence, this says that $V_1, V_2$ are independent conditioned on
  $V_1 \cap V_2$.
\end{lemma}

Notice that $h'$ can be written equivalently as
$h'(U) = h_1(U\cap V_1)+h_2(U\cap V_2) - h(U\cap Z)$.  While each term
is a normal polymatroid, it is not obvious that $h'$ is too, because
of the difference operation.  In fact, if $h_1, h_2, h$ are
polymatroids, then $h'$ is not a polymatroid in general.

The two lemmas prove Theorem~\ref{th:lpbase:eq:lptdb}, by showing that
$b_{\text{Berge}}\leq b_{\text{base}}$, as follows.  Consider any
feasible solution $h$ to \lptdb.  Consider first a single atom
$R_j(V_j)$ of $Q$: $h$ is only defined on all its variables and on the
entire set $V_j$.  By Lemma~\ref{lemma:normal:extension:of:one:bag},
we can extend $h$ to a normal polymatroid
$h: 2^{V_j} \rightarrow \Rp$.  We do this separately for each
$j=1,m$.  Next, we stitch these polymatroids together in order to
construct a polymatroid on all variables,
$h:2^{\set{X_1,\ldots, X_n}}\rightarrow \Rp$, and for this purpose we
use the Stitching Lemma~\ref{lemma:stich}.  Notice that the Lemma is
stronger than what we need, since in our case the intersection
$V_1 \cap V_2$ always has size 1 (since $Q$ is Berge-acyclic): we need
the stronger version for our third algorithm described in Sec.~\ref{subsec:lptd}.  By
using the conditional independence equality~\eqref{eq:independent}, we
can prove that $E_Q = h(X_1\cdots X_n)$, which completes the proof of
Theorem~\ref{th:lpbase:eq:lptdb}.

It remains to prove the two lemmas.

\begin{proof}[Proof of Lemma~\ref{lemma:normal:extension:of:one:bag}]
  We briefly review an alternative definition of normal polymatroids
  from~\cite{DBLP:conf/lics/Suciu23}.  For any $U \subseteq V$, the
  step function at $U$ is $h^U$ defined as:
\begin{align}
\forall X \subseteq V: &&  h^U(X) \defeq &
                  \begin{cases}
                    1 & \mbox{if $U\cap X\neq \emptyset$}\\
                    0 & \mbox{otherwise}
                  \end{cases} \label{eq:step:function}
\end{align}
When $U=\emptyset$, then $h^U \equiv 0$, so we will assume
w.l.o.g. that $U \neq \emptyset$.  A function $h : 2^V \rightarrow \R$
is a normal polymatroid iff it is a non-negative linear combination of
step functions:
\begin{align}
  h = & \sum_{U \subseteq V, U \neq \emptyset} c_U h^U
\end{align}
where $c_U \geq 0$ for all $U$.

We prove the lemma by induction on $n$, the number of variables in
$V$.  If $n=1$ then the lemma holds trivially because we define
$h(X_1) \defeq a_1$, so assume $n \geq 2$.  Rename variables such that
$a_1 \geq a_2 \geq \cdots \geq a_n$, and let $k \leq n$ be the
smallest number such that $a_1 + \cdots + a_k \geq A$: such $k$ must
exist by assumption of the lemma.  We prove the lemma in two cases.

{\bf Case 1}: $k=n$.  Let $\delta \defeq \sum_{i=1,n}a_i - A$ be the
excess of the inequality~\eqref{eq:a:a}: notice that $a_1 \geq \delta$
and $a_n \geq \delta$.  We define $h$ as follows:
\begin{align*}
  h = & (a_1-\delta)h^{X_1}+\delta h^{X_1,X_n}+\sum_{i=2,n-1}a_i h^{X_i} + (a_n-\delta)h^{X_n}
\end{align*}
By construction, $h$ is a normal polymatroid, and one can check by
direct calculation that $h(X_i)=a_i$ for all $i$ and $h(V)=A$.

{\bf Case 2}: $k < n$.  We prove by induction on $m=k,k+1,\ldots,n$
that there exists a normal polymatroid
$h : 2^{\set{X_1, \ldots, X_m}} \rightarrow \R$ s.t. $h(X_i)=a_i$ for
$i=1,m$ and $h(X_1\ldots X_m)=A$.  The claim holds for $m=k$ by Case
1.  Assuming it holds for $m-1$, let
$h' : 2^{\set{X_1,\ldots,X_{m-1}}}\rightarrow \R$ be such that
$h'(X_i)=a_i$ for $i=1,m-1$ and $h'(X_1\cdots X_{m-1})=A$.  We show
that we can extend it to $X_m$.  For that we first represent $h'$ over
the basis of step functions:
\begin{align*}
  h' = & \sum_{U \subseteq \set{X_1,\ldots,X_{m-1}}, U\neq \emptyset}c_U  h^U
\end{align*}
for some coefficients $c_U \geq 0$, and note that
$\sum_U c_U = h'(X_1\ldots X_{m-1})=A$.  Define $h$ as follows:
  \begin{align*}
    h = & \sum_{U \subseteq \set{X_1,\ldots,X_{m-1}}, U\neq \emptyset}c_U\left(1-\frac{a_m}{A}\right) h^U+ \sum_{U \subseteq \set{X_1,\ldots,X_{m-1}}, U\neq \emptyset}c_U\frac{a_m}{A} h^{U\cup\set{X_m}}
  \end{align*}
  By assumption of the lemma $a_m \leq A$, which implies that all
  coefficients above are $\geq 0$, hence $h$ is a normal
  polymatroid. Furthermore, by direct calculations we check that, for
  $i<m$, $h(X_i) = h'(X_i) = a_i$ (because
  $h^{U\cup \set{X_m}}(X_i)=h^U(X_i)$), and, for $i<m$,
  $h(X_m) = a_m$, because $h^U(X_m) = 0$ and
  $h^{U\cup\set{X_m}}(X_m)=1$, and the claim follows from
  $\sum_U c_U = h'(X_1\cdots X_{m-1})=A$.  Finally, we also have
  $h(X_1\ldots X_m) = \sum_U c_U = A$, as required.
\end{proof}

Finally, we prove the Stitching Lemma~\ref{lemma:stich}.  For that we
need two propositions.

\begin{proposition} \label{prop:technical:1} Let $h : 2^V \rightarrow \R$
  be a normal polymatroid, and $V_0 \supseteq V$ a superset of
  variables.  Define $h': 2^{V_0}\rightarrow \R$ by
  $h'(U) \defeq h(U \cap V)$ for all $U \subseteq V_0$.  Then $h'$ is
  a normal polymatroid.  In other words, $h'$ extends $h$ to $V_0$ by
  simply ignoring the additional variables.
\end{proposition}

\begin{proof}
  We verify condition~\eqref{eq:normal} directly.  When
  $U \subseteq V$, then $h'(W)=h(W)$ for all $W \subseteq U$ and the
  condition holds because $h$ is a normal polymatroid.  When
  $U\not\subseteq V$, then we claim that
  $\sum_{W\subseteq U}(-1)^{|W|+1}h(W \cap V)=0$.  Indeed, fix a variable
  $X_i \in U$, $X_i \not\in V$, and pair every subset $W \subseteq U$
  that does not contain $X_i$ with $W' \defeq W \cup \set{X_i}$.  Then
  $h(W\cap V)=h(W'\cap V)$ and the two terms corresponding to $W$ and
  $W'$ in~\eqref{eq:normal} cancel out, proving that the
  expression~\eqref{eq:normal} is $=0$.
\end{proof}

\begin{proposition} \label{prop:technical:2} Let $h: 2^V \rightarrow \R$ be
  a normal polymatroid, and $Z\subseteq V$ a subset of variables.
  Define the following set functions $h', h'' : 2^V\rightarrow \R$:
  \begin{align}
\forall U \subseteq V: &&    h'(U) \defeq & h(U\cap Z) &  h''(U) \defeq & h(U|U\cap Z) \label{eq:hprime:normal}
  \end{align}
  (Recall that $h(B|A) = h(AB)-h(A)$.)  Then both $h', h''$ are normal
  polymatroids.
\end{proposition}

\begin{proof}
  We consider two cases as above.  When $U\subseteq Z$, then for all
  $W \subseteq U$ we have $h'(W)=h(W)$, and $h''(W)=0$:
  condition~\eqref{eq:normal} holds for $h'$ because it holds for $h$,
  and it holds for $h''$ trivially since it is $=0$.  No consider
  $U\not\subseteq Z$.  Then we claim that the
  expression~\eqref{eq:normal} for $h'$ is $0$:
  \begin{align*}
    \sum_{W \subseteq U} (-1)^{|W|+1}h'(W)=& \sum_{W \subseteq U} (-1)^{|W|+1}h(W\cap Z)=0
  \end{align*}
  We use the same argument as in the previous lemma: pick a variable
  $X_i$ s.t. $X_i \in U$ and $X_i \not\in Z$ and pair each set
  $W \subseteq U$ that does not contain $X_i$ with the set
  $W' \defeq W \cup \set{X_i}$.  Then $h(W\cap Z)=h(W'\cap Z)$ and two
  terms for $W$ and $W'$ cancel out.  Finally,
  condition~\eqref{eq:normal} for $h''$ follows similarly:
  \begin{align}
    \sum_{W\subseteq U}(-1)^{|W|+1}h''(W) = & \sum_{W\subseteq U}(-1)^{|W|+1}h(W|W\cap Z) =\underbrace{\sum_{W\subseteq U}(-1)^{|W|+1}h(W)}_{\geq 0}-\underbrace{\sum_{W\subseteq U}(-1)^{|W|+1}h(W\cap Z)}_{=0}
  \end{align}
  The first term is $\geq 0$ because $h$ is a normal polymatroid, and
  the second term is $=0$, as we have seen.
\end{proof}

Finally, we prove the Stitching Lemma~\ref{lemma:stich}.

\begin{proof}[Proof of Lemma~\ref{lemma:stich}] We first use the two
  propositions to show that $h'$ from Eq.~\eqref{eq:stich} is a normal polymatroid.  Define two
  helper functions $h'_1 : 2^{V_1} \rightarrow \R$ and
  $h'_2 : 2^{V_2} \rightarrow \R$:
  \begin{align*}
\forall U \subseteq V_1:\   h'_1(U) \defeq & h_1(U|U \cap Z) & 
\forall U \subseteq V_2:\   h'_2(U) \defeq & h_2(U|U \cap Z)
  \end{align*}
  By Lemma~\ref{prop:technical:2}, both $h'_1, h'_2$ are normal
  polymatroids.  Next, we extend $h'_1, h'_2, h$ to the entire set
  $V_1 \cup V_2$ by defining:
  \begin{align*}
  \forall U \subseteq V_1 \cup V_2:&& h''_1(U) \defeq &h'_1(U\cap V_1)  & h''_2(U) \defeq &h'_2(U\cap V_2)  & h''(U) \defeq & h(U\cap Z)
  \end{align*}
  By Lemma~\ref{prop:technical:1} each of them is a normal
  polymatroid.  Since $h'$ in the corollary is their sum, it is also a
  normal polymatroid.

  We check that it agrees with $h_1$ on $V_1$.  For any $U \subseteq
  V_1$, we have $h_2(U\cap V_2|U \cap Z) = 0$ therefore:
  \begin{align*}
    h'(U) = & h_1(U \cap V_1|U \cap Z) + h(U \cap Z)= h_1(U|U\cap Z)+h_1(U\cap Z) = h_1(U)
  \end{align*}
  Similarly, $h'$ agrees with $h_2$ on $V_2$.  Finally,
  condition~\eqref{eq:independent} follows by setting
  $U:= V_1 \cup V_2$ in~\eqref{eq:stich} and applying the definition
  of conditional: $h(B|A)=h(B)-h(A)$ when $A \subseteq B$.
\end{proof}

\subsection{\lptd: Using Hypertree Decomposition}
\label{subsec:lptd}

The \lptdb algorithm is strictly limited by two requirements: $Q$
needs to be Berge-acyclic, and all statistics need to be full.  We
describe here a generalization of \lptdb, called \lptd, which drops
these two limitations.  When the restrictions of \lptdb are met, then
\lptd is slightly less efficient, however, its advantage is that it
can work on any query and constraints, without any restrictions.

A \emph{Hypertree Decomposition} of a full conjunctive query $Q$
 is a pair $(T,\chi)$, where $T$ is a tree and $\chi:
 \nodes(T)\rightarrow 2^V$ satisfying the following:
 \begin{itemize}
 \item For every variable $X_i$, the set
   $\setof{t \in \nodes(T)}{X_i \in \chi(t)}$ is connected.  This is
   called the \emph{running intersection property}.
 \item For every atom $R_j(V_j)$ of $Q$, $\exists t \in \nodes(T)$
   s.t. $V_j \subseteq \chi(t)$.
 \end{itemize}
 Each set $\chi(t) \subseteq V$ is called a \emph{bag}.  The
 \emph{width} of the tree $T$ is defined as
 $w(T) \defeq \max_{t \in \nodes(T)}|\chi(t)|$. We review a lemma by
 Lee~\cite{DBLP:journals/tse/Lee87}:

 \begin{lemma} \label{lemma:tony:lee} ~\cite{DBLP:journals/tse/Lee87}
   Let $(T,\chi:\nodes(T)\rightarrow 2^V)$ have the running
   intersection property and let $h : 2^V \rightarrow \R$ be a set
   function.  Define:
  \begin{align}
    E_{T,h} \defeq & \sum_{t \in \nodes(T)}h(\chi(t))-\sum_{(t_1,t_2)\in\edges(T)}h(\chi(t_1)\cap\chi(t_2)) \label{eq:et}
  \end{align}
  (1) If $h$ is a polymatroid (i.e. it satisfies the basic Shannon
  inequalities), then $E_{T,h} \geq h(V)$. (2) Suppose $h$ is the
  entropic vector defined by a uniform probability distribution on a
  relation $R(X_1, \ldots, X_n)$.  Then, $E_{T,h}=h(V)$ if for every
  $(t_1,t_2)\in \edges(T)$, the join dependency
  $R = \Pi_{V_1}(R) \Join \Pi_{V_2}(R)$ holds, where
  $V_1, V_2\subseteq V$ are the variables occurring on the two
  connected components of $T$ obtained by removing the edge
  $(t_1,t_2)$.
\end{lemma}

For a simple illustration, consider the 3-way join $J_3$
(Eq.~\eqref{eq:j3}).  Its tree decomposition $T$ has 3 bags
$XY, YZ, ZU$, and $E_{T,h} = h(XY)+h(YZ)+h(ZU)-h(Y)-h(Z)$; one can
check that $E_{T,h} \geq h(XYZU)$ using two applications of
submodularity.

Our new linear program, called \lptd, is constructed from a
hypertree decomposition $(T,\chi)$ of the query as follows:

\smallskip

\noindent {\bf The Real-valued Variables} are all expressions of the
form $h(U)$ for $U \subseteq \chi(t)$, $t \in \nodes(T)$.  The total
number of real-valued variable is $\sum_{t \in \nodes(T)}
2^{|\chi(t)|}$, i.e. it is exponential in the tree-width of the query.

\smallskip

\noindent {\bf The Objective} is to maximize $E_{T,h}$
(Eq.~\eqref{eq:et}), subject to the following constraints.

\smallskip

\noindent {\bf Statistics Constraints:}
All statistics constraints Eq.~\eqref{eq:h:p} of \lpbase.
Since we don't have numerical variable $h(U)$ for all $U$, we must
check that~\eqref{eq:h:p} uses only available numerical
variables.  This holds, because each statistics refers to some atom
$R_j(V_j)$, and there exists of some bag such that
$V_j \subseteq \chi(t)$, therefore we have numerical variables $h(U)$
for all $U \subseteq V_j$.

\smallskip

\noindent {\bf Normality Constraints:}
For each bag $\chi(t)$, \lptd contains all constraints of the
form~\eqref{eq:normal}.  In other words, the restriction of $h$ to
$\chi(t)$ is normal.

We prove:

\begin{theorem} \label{th:lptd} \lpbase and \lptd compute the same
  value.
\end{theorem}

The theorem holds only when all degree constraints used in the
statistics are \emph{simple}, as we assumed throughout this paper.
For a simple illustration, the \lptd for $J_3$ consists of 7 numerical
variables \\
$h(X),h(Y),h(Z),h(U),h(XY),h(YZ),h(ZU)$ (we omit
$h(\emptyset)=0$) and the following Normality Constraints:
\begin{align*}
  h(X)+h(Y)-h(XY)\geq & 0 & h(Y)+h(Z)-h(YZ)\geq & 0 \\
  h(Z)+h(U)-h(ZU) \geq & 0
\end{align*}

To compare \lptd and \lptdb, assume that the query $Q$ is Berge-acyclic
and all statistics are on simple and full degree constraints.  The
difference is that, for each atom $R_j(V_j)$, \lptdb has only $1+|V_j|$
real-valued variables and only $1+|V_j|$ additivity constraints, while
\lptdb has $2^{|V_j|}$ variables and normality constraints.

In the remainder of this section we prove Theorem~\ref{th:lptd}.

Denote by $b_{\text{base}}$ and $b_{\text{td}}$ the optimal solutions
of \lpbase and \lptd respectively.  We will prove that
$b_{\text{base}}=b_{\text{td}}$.

First, we claim that $b_{\text{base}}\leq b_{\text{td}}$.  It is known
from~\cite{DBLP:journals/pacmmod/KhamisNOS24} that \lpbase has an
optimal solution $h^*$ that is a normal polymatroid; thus
$b_{\text{base}}=h^*(V)$ (recall that $V$ is the set of all
variables), where $h^*$ is normal.  Then $h^*$ is also a feasible
solution to \lptd, therefore its optimal value is at least as large as
the value given by $h^*$, in other words
$b_{\text{td}}\geq E_{T,h^*}$.  By Lemma~\ref{lemma:tony:lee}, we have
$E_{T,h^*}\geq h^*(V) = b_{\text{base}}$, which proves the claim.

Second, we prove that $b_{\text{td}}\leq b_{\text{base}}$ by using the Stitching Lemma~\ref{lemma:stich}.
Let $h^*$ be an optimal solution to \lptd, thus
$b_{\text{td}}=E_{T,h^*}$.  The function $h^*$ is defined only on
subsets of the bags $\chi(t)$, $t \in \nodes(T)$, and on each such
subset, it is a normal polymatroid.  We extend it to a normal
polymatroid defined on all variables
$V = \bigcup_{t \in \nodes(T)}\chi(t)$ by repeatedly applying the
Stitching Lemma~\ref{lemma:stich}. Condition~\eqref{eq:independent} of
the corollary implies that this extension satisfies
$h^*(V)=E_{T,h^*}$.  Thus, $h^*$ is a normal polymatroid, and, hence,
a feasible solution to \lpbase.  It follows that the optimal value of
\lpbase is at least as large as that given by $h^*$, in other words
$b_{\text{base}} \geq h^*(V)$.  This completes the proof of the claim.


\subsection{\lpflow: Missing Details from Section~\ref{subsec:lpflow}}
\label{subsec:lpflow:details}

Section~\ref{subsec:lpflow} gives the high-level idea of the \lpflow algorithm
using an example. We give here a more formal description of the algorithm
and prove Theorem~\ref{th:lpbase:eq:lpflow}.
The input to \lpflow is an arbitrary conjunctive query $Q$ of the form Eq.~\eqref{eq:cq}
(not necessarily a full query) and a set of statistics on the input database consisting of $\ell_p$-norms
of {\em simple} degree sequences, i.e. statistics of the form $\lp{\degree_{R_j}(V|U)}_p$ where $|U|\leq 1$.
For the purpose of describing \lpflow,
we construct a flow network $G=(\nodes,\edges)$ that is defined as follows:
(Recall that $\vars(Q) =\{X_1, \ldots, X_n\}$ is the set of variables of the query.)
\begin{itemize}
    \item The set of nodes $\nodes\subseteq 2^{\vars(Q)}$ consists of the following nodes:
    \begin{itemize}
        \item The node $\emptyset$, which is the source node of the flow network.
        \item A node $\{X_i\}$ for every variable $X_i \in \vars(Q)$.
        \item A node $UV$ for every statistics $\lp{\degree_{R_j}(V|U)}_p$. 
    \end{itemize}
    \item The set of edges $\edges$ consists of two types of edges:
    \begin{itemize}
        \item {\bf Forward edges:} These are edges of the form $(a, b)$
        where $a, b \in \nodes$ and $a \subset b$. Each such edge $(a, b)$ has a finite capacity $c_{a, b}$. In particular, for every statistics
        $\lp{\degree_{R_j}(V|U)}_p$, we have two forward edges:
        One edge $(\emptyset, U)$ and another $(U, UV)$. (Recall that $|U|\leq 1$.)
        \item {\bf Backward edges:} These are edges of the form $(a, b)$
        where $a, b \in \nodes$ and $b \subset a$. Each such edge $(a, b)$ has an infinite
        capacity $\infty$.
        In particular, for every statistics $\lp{\degree_{R_j}(V|U)}_p$ and
        every variable $X_i \in UV$, we have a backward edge $(UV, \{X_i\})$.
    \end{itemize}
\end{itemize}
We are now ready to describe the linear program for \lpflow. Recall that $V_0$ is the set of \groupby variables in the query $Q$ from Eq.~\eqref{eq:cq}.

\smallskip

\noindent {\bf The Real-valued Variables} are of two types:
\begin{itemize}
    \item Every statistics $\lp{\degree_{R_j}(V|U)}_p$ has an associated {\em non-negative}
    variable $w_{U,V,p}$.
    \item For every \groupby variable $X_i \in V_0$, we have a flow variable $f_{a, b; X_i}$
    for every edge $(a, b) \in \edges$.
\end{itemize}

\smallskip

\noindent {\bf The Objective} is to minimize the following sum over all available statistics
$\lp{\degree_{R_j}(V|U)}_p$:
\begin{align}
    \sum w_{U,V,p} \cdot \log \lp{\degree_{R_j}(V|U)}_p
    \label{eq:lpflow:objective}
\end{align}

\smallskip

\noindent {\bf The Constraints} are of two types:
\begin{itemize}
    \item {\em Flow conservation constraints:} For every \groupby variable $X_i \in V_0$,
    the variables $f_{a, b;X_i}$ must define a valid flow from the source node
    $\emptyset$ to the sink node $\{X_i\}$ that has a capacity $\geq 1$.
    This means that for every node $c \in \nodes - \{\emptyset\}$, we must have:
    \begin{align}
        \sum_a f_{a, c;X_i} -\sum_b f_{c, b; X_i} \geq 1, & \quad\quad\text{ if $c =\{X_i\}$}\label{eq:flow:conservation:1}\\
        \sum_a f_{a, c;X_i} -\sum_b f_{c, b; X_i} \geq 0, & \quad\quad\text{ otherwise}
        \label{eq:flow:conservation:2}
    \end{align}
    \item {\em Flow capacity constraints:} For every \groupby variable $X_i \in V_0$
    and every {\em forward} edge $(a, b)$, the flow variable $f_{a, b;X_i}$ must satisfy:
    \begin{align}
        f_{a, b;X_i} \leq c_{a, b} \label{eq:flow:capacity}
    \end{align}
    where $c_{a, b}$ is the {\em capacity} of the forward edge $(a, b)$.
    (Recall that backward edges have infinite capacity.)
    The capacity variables $c_{a, b}$ are determined by the statistics constraints.
    In particular, every statistics $\lp{\degree_{R_j}(V|U)}_p$ contributes a capacity
    of $w_{U,V,p}$ to $c_{U,UV}$ and a capacity of $\frac{w_{U,V,p}}{p}$ to $c_{\emptyset,U}$.
    Formally,
    \begin{align}
        c_{\emptyset,U} &\defeq
        \sum_{p} w_{\emptyset,U,p}
        +\sum_{V,p} \frac{w_{U,V,p}}{p}\label{eq:lpflow:c}\\
        c_{U, UV} &\defeq
        \sum_{p} w_{U,V,p} &\text{ if $U \neq \emptyset$}\nonumber
    \end{align}
\end{itemize}
\smallskip

We are now ready to prove Theorem~\ref{th:lpbase:eq:lpflow}, which says that the linear programs
for \lpbase and \lpflow have the same optimal value.
To that end, we first write the dual LP for \lpbase.
For every statistics constraint of the form Eq.~\eqref{eq:h:p}, we introduce a dual variable $w_{U,V,p}$. The dual of \lpbase is equivalent to:
\begin{align}
    \min\quad &\sum w_{U,V,p} \cdot \log \lp{\degree_R(V|U)}_p\label{eq:lpbase:dual}\\
    \text{s.t.}\quad& \text{The following is a valid Shannon inequality:}\nonumber\\
    &h(V_0) \leq \sum w_{U,V,p} \left(\frac{1}{p}h(U)+h(V|U)\right)\label{eq:lpflow:shannon}\\
    & w_{U,V,p} \geq 0\nonumber
\end{align}
Inequality~\eqref{eq:lpflow:shannon} satisfies the property that for every $h(V|U)$
on the RHS, we have $|U| \leq 1$. In order to check that such an inequality is a valid
Shannon inequality, we rely on a key result from~\cite{DBLP:journals/corr/abs-2211-08381}.
In particular,~\cite{DBLP:journals/corr/abs-2211-08381} is concerned
with Shannon inequalities of the following form.
Let $\calX=\{X_1, \ldots, X_n\}$ be a set of variables, and $\calC$
be a set of distinct pairs $(U, V)$ where $U, V\subseteq \calX$, $U \cap V = \emptyset$
and $|U| \leq 1$.
For every pair $(U,V)\in\calC$, let $c_{U,UV}$ be a non-negative constant.
Moreover, let $V_0$ be a subset of $\calX$.
Consider the following inequality:
\begin{align}
    h(V_0) \leq \sum_{(U,V)\in\calC} c_{U, UV} h(V|U)
    \label{eq:lpflow:shannon:general}
\end{align}
\cite{DBLP:journals/corr/abs-2211-08381} describes a reduction from
the problem of checking whether Eq.~\eqref{eq:lpflow:shannon:general} is a valid Shannon inequality to a collection of $|V_0|$ network flow problems.
These flow problems are over the same network $G=(\nodes, \edges)$,
which is similar to the flow network described above for \lpflow. In particular,
\begin{itemize}
    \item The nodes are $\emptyset$, $\{X_i\}$ for each variable $X_i \in \calX$, and $UV$ for each $(U,V)\in\calC$.
    \item The edges have two types:
    \begin{itemize}
        \item Forward edges: For each $(U,V)\in\calC$, we have a forward edge $(U,UV)$ with capacity $c_{U,UV}$.
        \item Backward edges: For each $(U,V)\in\calC$ and each variable $X_i \in UV$,
        we have a backward edge $(UV, \{X_i\})$ with infinite capacity.
    \end{itemize}
\end{itemize}
\begin{lemma}[\cite{DBLP:journals/corr/abs-2211-08381}]
    Inequality~\eqref{eq:lpflow:shannon:general} is a valid Shannon inequality if and only if
    for each variable $X_i \in V_0$, there exists a flow $\left(f_{a, b;X_i}\right)_{(a, b)\in\edges}$ from the source node $\emptyset$
    to the sink node $\{X_i\}$ with capacity at least $1$.
    In particular, the flow variables $\left(f_{a, b;X_i}\right)_{(a, b)\in\edges}$ must satisfy the flow conservation constraints~\eqref{eq:flow:conservation:1} and~\eqref{eq:flow:conservation:2} and the flow capacity constraints~\eqref{eq:flow:capacity}.
\end{lemma}
Using the above lemma, we can prove Theorem~\ref{th:lpbase:eq:lpflow} as follows.
Take inequality~\eqref{eq:lpflow:shannon} and group together identical conditionals
on the RHS in order to convert it into the form of Eq.~\eqref{eq:lpflow:shannon:general}.
The coefficients $c_{U,UV}$ of the resulting Eq.~\eqref{eq:lpflow:shannon:general}
will be identical to those defined by Eq.~\eqref{eq:lpflow:c}.
Then, we can apply the lemma to check the validity of Eq.~\eqref{eq:lpflow:shannon}.
But now, the dual LP~\eqref{eq:lpbase:dual} for \lpbase is equivalent to the linear program for \lpflow.

\end{document}